\documentclass[%
reprint,
superscriptaddress,
nofootinbib,
amsmath,amssymb,
aps,
prd,
]{revtex4-1}
\usepackage{graphicx}
\usepackage{dcolumn}
\usepackage{bm}
\usepackage{hyperref}
\usepackage[dvipsnames]{xcolor}
\usepackage{multirow}

\definecolor{aquamarine}{rgb}{0.2,0.7,0.6}
\definecolor{cerulean}{RGB}{0,166,214} 
\definecolor{hypershade}{rgb}{0.3,0.3,0.8}
\definecolor{subtlered}{rgb}{0.8,0.3,0.3}

\newcommand{\Dsource}{D_{\rm S}}
\newcommand{\Dlens}{D_{\rm L}}
\newcommand{\Dls}{D_{\rm LS}}
\newcommand{\thetaE}{\theta_{\rm E}}
\def\tE{t_{\rm E}}
\def\rE{R_{\rm E}}
\def\uT{u_{\rm T}}

\usepackage{comment}

\hypersetup{
  colorlinks=true,
  citecolor=cerulean,
  urlcolor=aquamarine,
  linkcolor=cerulean
}
\usepackage{tikz}
\usepackage{scalerel}
\usetikzlibrary{svg.path}
\definecolor{orcidlogocol}{HTML}{A6CE39}
\tikzset{orcidlogo/.pic={
 \fill[orcidlogocol] svg{M256,128c0,70.7-57.3,128-128,128C57.3,256,0,198.7,0,128C0,57.3,57.3,0,128,0C198.7,0,256,57.3,256,128z};
 \fill[white] svg{M86.3,186.2H70.9V79.1h15.4v48.4V186.2z}
 svg{M108.9,79.1h41.6c39.6,0,57,28.3,57,53.6c0,27.5-21.5,53.6-56.8,53.6h-41.8V79.1z M124.3,172.4h24.5c34.9,0,42.9-26.5,42.9-39.7c0-21.5-13.7-39.7-43.7-39.7h-23.7V172.4z}
 svg{M88.7,56.8c0,5.5-4.5,10.1-10.1,10.1c-5.6,0-10.1-4.6-10.1-10.1c0-5.6,4.5-10.1,10.1-10.1C84.2,46.7,88.7,51.3,88.7,56.8z};
}}
\newcommand\orcidicon[1]{\href{https://orcid.org/#1}{\mbox{\scalerel*{
\begin{tikzpicture}[yscale=-1,transform shape]
\pic{orcidlogo};
\end{tikzpicture}
}{|}}}}

\newcommand{\apjl}{apjl}
\newcommand{\aap}{aap}

\begin{document}

\title{Microlensing of fast and slow compact objects}

\author{Manish Tamta$^{\orcidicon{0009-0006-7088-8705}}$}
\email{manishtamta@iisc.ac.in}
\affiliation{Centre for High Energy Physics, Indian Institute of Science, C. V. Raman Avenue, Bengaluru 560012, India}%

\author{Nirmal Raj$^{\orcidicon{0000-0002-4378-1201}}$}
\email{nraj@iisc.ac.in}
\affiliation{Centre for High Energy Physics, Indian Institute of Science, C. V. Raman Avenue, Bengaluru 560012, India}%

\author{Himanshu Verma$^{\orcidicon{0000-0002-6302-251X}}$}
\email{hverma@lsu.edu}
\affiliation{Department of Physics and Astronomy, Louisiana State University, Baton Rouge, LA 70803, USA}%

\begin{abstract}
Gravitational microlensing constraints on non-standard compact objects are conventionally derived assuming lenses trace the dark matter halo with velocities following a Maxwell-Boltzmann distribution centered around $10^{-3}c$.
However, a variety of theoretical scenarios predict populations of compact objects 
whose velocities deviate dramatically from those of virialized halo dark matter -- 
ultrarelativistic primordial black holes from cosmic string collapse, mirror neutron stars, gravitationally kicked black hole merger remnants, dark matter nuggets, free floaters ejected from gravitationally bound systems, disk-formed compact objects, and so on.
For a given Einstein crossing time, the speed-mass degeneracy inherent in it means that fast (slow) lenses produce events at larger (smaller) masses than spanned by standard windows, opening qualitatively new regions of parameter space. 
After deriving model-independent upper limits on the microlensing event rate, we obtain mass-dependent constraints on the density of lens populations with speeds spanning $10^{-4}c-10^{-1}c$ from surveys of M31 by Subaru-HSC and the LMC by OGLE with different observing cadences.
We do this for two benchmark velocity distributions -- Maxwell-Boltzmann and Dirac delta -- and two spatial distributions -- uniform and NFW, and exclude lens densities and masses that differ from dark matter constraints by orders of magnitude.
We examine the effect of the transverse motion of the source and observer relative to the lensing tube, which becomes significant for our slow lenses. 
We also show that, unlike in dark matter searches, for our fast lenses an increase in the cadence of observations would probe smaller masses without suppression of event rates from the finite source and wave optics effects.
\end{abstract}

\maketitle

\section{\label{sec:intro}Introduction}

Gravitational microlensing -- the temporary, achromatic magnification of a background star by a transiting body~\cite{Einstein:1936llh,NarayanBartelmann} -- has long served as a powerful, model-independent probe of compact objects along the line of sight to luminous sources. 
Surveys such as MACHO, EROS-2, OGLE, and Subaru-HSC, monitoring tens of million stars in the Magellanic Clouds, Galactic Bulge and M31, have placed stringent constraints on the abundance of primordial black holes (PBHs) and other massive compact halo objects (MACHOs) as constituents of dark matter,
which apply as well to extended dark objects such as microhalos and boson stars~\cite{reviewPBH:GreenKavanagh:2020jor,Green:2025dut,WidrowHClouds,Fairbairn:2017dmf,Blinov_2020,CroonMcKeenRajECOlocation2,CroonMcKeenRajECOlocation1,BaiDMACHOs:2020jfm,AnsariArunQBalls:2023cay}.
These constraints are conventionally derived under the assumption that the lens population traces the dark matter halo of the Milky Way and takes up a Maxwell-Boltzmann velocity distribution characterized by the local circular speed $\sim 10^{-3}c$. 
The foregrounds to these searches are faint celestial bodies, typically main sequence stars, brown dwarfs, and compact stellar relics. 
Upcoming surveys are also sensitive to populations of free-floating planets~\cite{FFPRoman:Sajadian2021,EuclidRomanjointreach:Bachelet2022}.

\begin{table*}[t]
\centering
\begin{tabular}{lllllll}
\toprule
survey & source field: $(\Dsource/{\rm kpc}, \ell, b)$ & $t_{\rm E, min}$ & $t_{\rm E, max}$ & $N_{\star}/10^6$ &  $N_{\rm excl} \ (N_{\rm obs})$ &\\
\hline
\vspace{.15cm}
Subaru-HSC~\cite{SubaruHSC:Niikura:2017zjd} & M31: (770, 121.2$^\circ$, $-$21.6$^\circ$)  & 2 min & 7 hours & 87 & 4.74 (1) &\\
OGLE-hc~\cite{OGLELMChc:Mroz:2024wia} & \multirow{4}{13em}{LMC: (50, 280.5$^\circ$, $-$31.2$^\circ$)}  & 16 min & 435 d & 35 & 4.74 (1) &\\
OGLE~\cite{OGLELMClc:Mroz:2024wag} &  & 1 d & 20 yr & 78.7 & 20.67 (13) &\\
\textcolor{gray}{EROS-2}~\cite{EROS2Tisserand} &  & \textcolor{gray}{1 d} & 
\textcolor{gray}{6.7 yr} & \textcolor{gray}{6.7} & \textcolor{gray}{4.74 (1)} &\\
\textcolor{gray}{MACHO}~\cite{MACHO:2000qbb} & & \textcolor{gray}{2 d} & \textcolor{gray}{5.7 yr} & \textcolor{gray}{11.9} & \textcolor{gray}{20.67 (13)} &\\
\hline
\end{tabular}
\caption{Particulars of the microlensing surveys used in this work. 
The last column shows the number of events expected at 95\% C.L. to set limits for the number of confirmed microlensing events observed, $N_{\rm obs}$. 
As seen in Fig.~\ref{fig:gammaexcl_vperpvsM} left panel, the Subaru-HSC and OGLE surveys yield the tightest limits, hence our main results in Fig.~\ref{fig:all_limits_nfw_uniform_MB_DD} use only those. 
See text for further details.}
\label{tab:surveydetails}
\end{table*}

Standard astrophysical bodies and macroscopic dark matter objects are, however, not the only objects that can produce microlensing signals. 
A variety of motivated astrophysical and particle physics scenarios predict populations of compact objects whose velocities deviate markedly from those of virialized halo dark matter. 
On the fast end, lenses with speeds $\gg 10^{-3}c$ cannot plausibly trace the local dark matter distribution and must instead be of extragalactic or exotic origin, especially if they exceed the Galactic escape velocity.
Candidate populations in this regime include ultrarelativistic PBHs produced from the collapse of cosmic strings~\cite{fastlenses:PBHcosmicstring:Jenkins:2020ctp}, mirror neutron stars endowed with natal kicks from mirror supernovae~\cite{fastlenses:mirrorNS:Hippert:2021fch}, (for comparison, standard neutron stars are observed to have kick speeds up to 1000~km/s~\cite{Disberg:2025xoh}; a mirror neutron star may attain greater kick speeds if a larger fraction of energy from its birthing supernova goes into its kinetic energy),
black hole remnants of binary mergers that receive gravitational ``superkicks'' of up to $\sim 5000~\mathrm{km/s}$~\cite{fastlenses:BHGWsuperkix:Gerosa:2016vip}, 
black holes ejected from dense stellar clusters via gravitational slingshot interactions~\cite{fastlenses:BHslingshot:Bamber:2025gxj}, 
and macroscopic dark matter nuggets accelerated by long-range fifth forces~\cite{fastlenses:longrangenuggets:Gresham:2022biw}. 
Mention must also be made of supermassive black holes gaining a velocity kick via gravitational wave recoil following a binary merger, an instance of which was recently confirmed by JWST observations~\cite{fastlenses:GWrecoilJWST:2025bah}.
One may think of these as macroscropic analogues of populations of ``boosted dark matter'' in particle contexts.

On the slow end, it is known that free-floating planets and their asteroid-equivalent inter-stellar objects inherit their speeds of ejection from their parent stellar system, typically $10^{-4}c$: their post-ejection velocities are weakly affected by gravitational scattering with stars as the stellar relaxation time is generally longer than their age~\cite{slowlenses:floaters:2026arXiv260212017L}.
Indeed, a free-floating planet was recently observed in microlensing with best-fit transverse speed $\mathcal{O}(10-100)$~km/s~\cite{slowlenses:FFPOGLE:2026Science}.
We can imagine similar mechanisms in a complex dark sector, with smaller structures coming unbound from larger ones.
In that regard, the aforesaid mirror neutron stars could also receive natal kick velocities $\ll 10^{-3}c$.
Objects ejected from a galaxy would generally slow down as they climb out of its gravitational potential.
Stellar disk-formed objects form with $\mathcal{O}(10^{-4})c$ dispersion speeds~\cite{slowlenses:diskformed1993}, and so can objects formed in ``dark disks"~\cite{slowlenses:darkdiskkaplinghat:2009ApJ}.
(The microlensing reach of Rubin to compact objects in a dark disk has been estimated~\cite{darkdisk:WinchSetfordCurtin:2020cju}.)
Further, we generically expect that dark matter with self-interacting, dissipative interactions would undergo fragmentation and gravitational collapse into compact objects~\cite{fragmented:Buckley:2017ttd,fragmented:DAmico:2017lqj, fragmented:ChangDanielEssigKouvaris:2018bgx,fragmented:Gurian:2022nbx,fragmented:LisantiCurtin:2023zar,fragmented:Bramante:2023ddr,fragmented:Osuna:2026dzj}; whether these structures would virialize with galactic halos punctually and follow dark matter velocity distributions is not obvious. 
All that said, we will show that microlensing has an important limitation that makes searches for these slow bodies generally challenging.
It arises from the ``bulk motion'' of the lensing tube -- the volume through which lens transits are counted as events -- arising from the transverse velocities of the source star and the observer (i.e., the Sun).
Since the speeds of stars in virialized halos are $\sim 10^{-3}c$, we can generically expect this effect to dominate the rate at which microlensing events occur for lenses with proper speeds $\ll 10^{-3}~c$.
We will nonetheless show results for the case of $10^{-4}c$ speed lenses, pretending the existence of a fantastic scenario: lenses overall co-moving with the lensing tube with occasional transits. 
While at the edge of plausibility, this picture is not impossible for objects in a co-rotating dark disk, or mirror neutron stars.
In any case, our results for slow lenses must be taken not so much as rigorous limits as an educative exercise.

\begin{figure*}[t]
    \centering
    \includegraphics[width=0.49\textwidth]{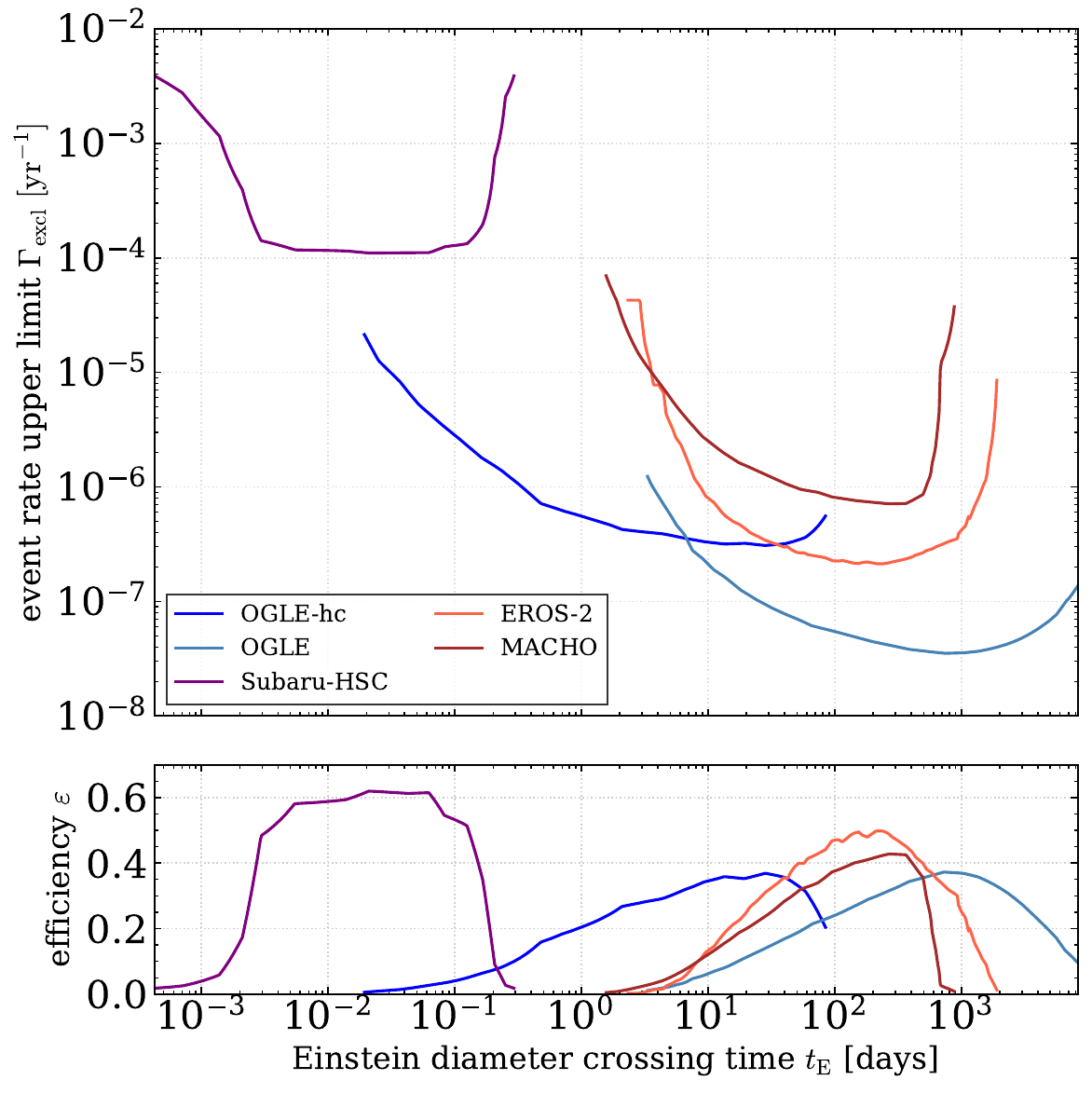}
     \includegraphics[width=0.49\textwidth]{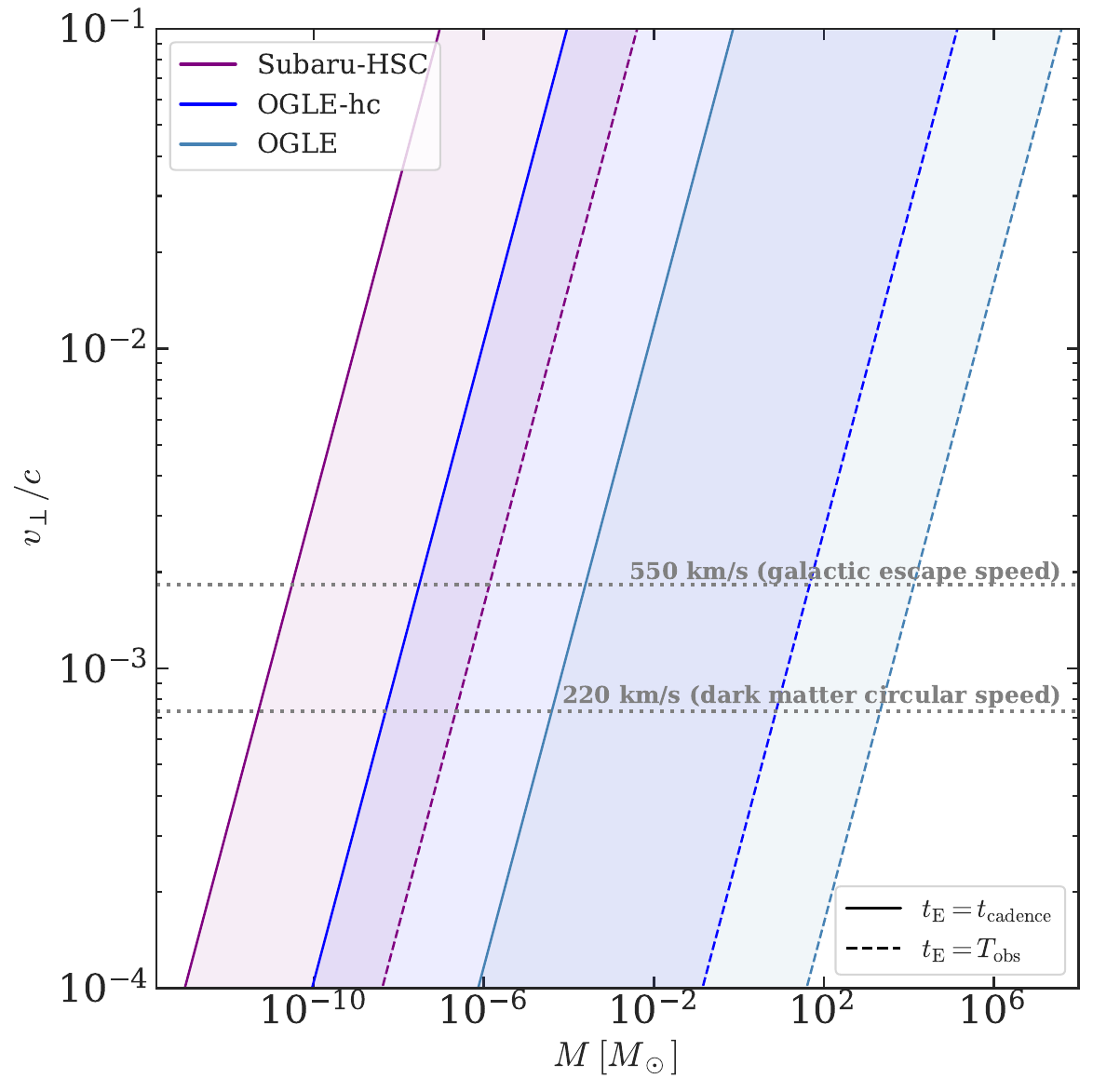}
    \caption{{\bf \em Left.} 95\%~C.L. upper limits on the microlensing event rate per source star as obtained from Eq.~\eqref{eq:Gamma_excl} and detection efficiencies taken from  the references in Table~\ref{tab:surveydetails} as a function of $\tE$ as defined in Eq.~\eqref{eq:tEvvperp} (differing by a factor of 2 from the $\tE$ of Ref.~\cite{EROS2Tisserand}, which we account for).
    For Subaru-HSC (OGLE-hc) [OGLE] we use the efficiencies for the median source brightness of 24 mag in Ref.~\cite{SubaruHSC:Niikura:2017zjd} (stellar size of $\rho = 1$ in Ref.~\cite{OGLELMChc:Mroz:2024wia}) [OGLE III+IV curve in Ref.~\cite{OGLELMClc:Mroz:2024wag}].  
   {\bf \em Right.} Rough estimate of ranges of lens speeds transverse to the lensing tube and lens masses probed by surveys given the minimum and maximum Einstein crossing times $\tE$ to which they are sensitive, as obtained from Eq.~\eqref{eq:tEvvperp}.
   See text for further details.}
    \label{fig:gammaexcl_vperpvsM}
\end{figure*}

The Einstein radius $\rE$ for a lens mass $M$ is proportional to $\sqrt{M}$ and is set by the lens-source geometry~\cite{NarayanBartelmann}; the duration of a microlensing event is then $\hat{t} \sim  \rE/v_\perp$, where $v_\perp$ is the lens velocity transverse to the line-of-sight.
Thus, standard analyses interpreting observations as dark matter objects, which effectively fix the velocity distribution, allows for mapping a range of $\hat{t}$ (to which the telescope is sensitive) onto a range of $M$, breaking what is fundamentally a two-parameter degeneracy. 
Allowing for other lens velocities re-introduces this degeneracy, thereby opening qualitatively new regions of the microlensing parameter space: for a given mass, fast- and slow-moving lenses would produce shorter- or longer-duration events, which may escape detection if they fall below cadence windows or last longer than observing baselines. 
A systematic exploration of microlensing signals across a broad range of lens velocities would therefore serve to provide the true discovery reach of photometric surveys.\footnote{Astrometric surveys can, however, break the lens mass-velocity degeneracy by measuring the microlensing parallax~\cite{parallaxgould1992ApJ,parallaxalcock1995ApJ,1995A&A...294..287H, 2016ApJ...830...41L,2017Univ....3...53L,2022ApJ...933...83S, 2022ApJ...933L..23L}.}

In this work, we derive microlensing limits for lenses that generically do not follow the phase space of dark matter.
First we show limits on the event rate from various microlensing surveys as a function of the Einstein diameter crossing time, which is the most generic interpretation of results at these searches.
We then estimate mass-dependent limits on the population of our slow and fast lenses for some benchmark distributions: velocities with a dispersion (Maxwell-Boltzmann distribution) and those that are uniform (Dirac delta distribution), and spatial distributions that are uniform and dark matter-like. 
Our results demonstrate that the microlensing discovery space is far richer than conventionally assumed, and that, due to a reduction of finite source and wave optics effects, dedicated searches for anomalously short events can probe exotic compact object populations that would otherwise evade standard analyses.
Our work is similar in spirit to some recent studies: Ref.~\cite{Green:2025dut} explored uncertainties in microlensing event rates from assumptions on dark matter density profiles and Maxwell-Boltzmann circular velocities, and Ref.~\cite{fragmented:Osuna:2026dzj} outlined microlensing signals for dark matter substructure with spatial and mass distributions governed by dissipative dynamics.
In contrast, we consider lens velocities spanning orders of magnitude and a flat density distribution, which scarcely mimic dark matter.

This paper is laid out as follows.
In Sec.~\ref{sec:setup} we present limits on the microlensing rate as a function of the Einstein crossing time, independent of lens masses and velocities, and the approximate regions probed by various observatories in the transverse speed vs mass space.
We then describe the lens spatial and velocity distributions considered in this work, and derive event rates for them.
In Sec.~\ref{sec:results} we show and describe limits on the populations of lenses for various benchmark speeds.
In Sec.~\ref{sec:discs} we discuss the future scope of our work and conclude.

\begin{figure*}[t]
    \centering    \includegraphics[width=0.99\textwidth]{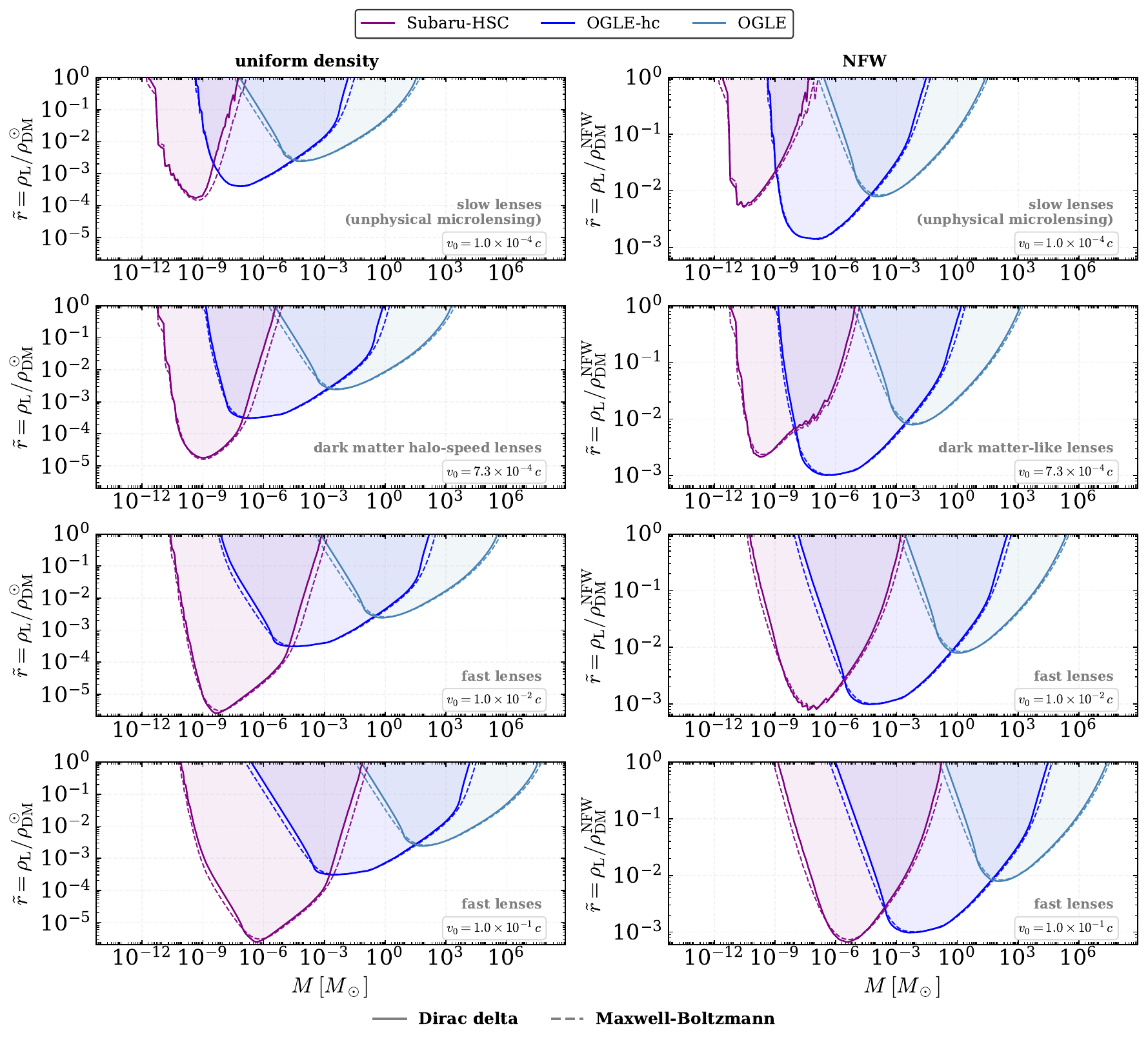}
    \caption{95\% C.L. limits as a function of lens mass $M$ on the ratio of our lens populations to benchmark dark matter populations as defined in Eq.~\eqref{eq:rhonorm}.
    The {\bf \em left} panels are for a spatial distribution of lenses that is uniform and the {\bf \em right} panels for that following a dark matter-like (NFW) profile.
    A given survey probes heavier (lighter) lenses if they are moving faster (slower) as illustrated in the right panel of Fig.~\ref{fig:gammaexcl_vperpvsM}.
    The top panels with $10^{-4}c$ speeds does not represent true limits for lenses with isotropic velocities as the event rate would be set by the bulk motion of the lensing tube with speeds $\sim10^{-3}c$; we display them here mainly to illustrate the trends of our limits.
    See Sec.~\ref{sec:results} for further details.}
    \label{fig:all_limits_nfw_uniform_MB_DD}
\end{figure*}

\section{Microlensing events}
\label{sec:setup}

\subsection{Model-independent limits}

In Fig.~\ref{fig:gammaexcl_vperpvsM} left top panel we show generic limits on the rate of microlensing events per source star from various surveys with no assumptions about the lenses' masses and speeds, obtained from
\begin{equation}
        \Gamma_\mathrm{excl}(\tE) =
    \frac{N_\mathrm{excl}}{N_\star\, T_\mathrm{obs}\, \varepsilon(\tE)}~,
    \label{eq:Gamma_excl}
    \end{equation}
where $N_{\rm excl}$ is the number of expected events at 95\%~C.L. for the number of observed events at a survey, $N_\star$ is the number of source stars in the star field used in the survey,
$T_{\rm obs}$ is the total observation time, and 
$\varepsilon(\tE)$ is the detection efficiency as a function of Einstein diameter crossing time, which is estimated by search collaborations by simulations and fits to the microlensing light curve; in the left bottom panel of Fig.~\ref{fig:gammaexcl_vperpvsM} we show the efficiencies used in this work.
In Table~\ref{tab:surveydetails} we give $N_{\rm excl}$, $N_\star$ and $T_{\rm obs}$ of various surveys, among other details.
We note that our limits are conservative as we fix $N_{\rm excl}$ to a constant value across the entire range of $\tE$, whereas the events observed at surveys confirmed to trace microlensing light-curves occur at some fixed $\tE$ within errors; outside such $\tE$ windows, we expect $N_{\rm excl}$ to be actually smaller.

Now we have 
\begin{equation}
\tE = \frac{2 \uT \rE}{v_\perp} = \frac{4\uT}{v_\perp}\sqrt{\frac{GM}{c^2} \frac{\Dlens \Dls}{\Dsource}}~,
\label{eq:tEvvperp}
\end{equation}
where \{$\Dlens, \Dls,\Dsource$\} are the \{lens-observer, lens-source, source-observer\} distances, and $\uT$ the impact parameter (in units of $\rE$) below which a lens transit produces a magnification above some detection threshold.
In this work, we take this threshold as 1.34 as done in the surveys we consider, which for point-like lenses makes $\uT = 1$ for point-like source stars~\cite{NarayanBartelmann}, but $\uT < 1$ when finite source effects are accounted for (see below). 
We can use Eq.~\eqref{eq:tEvvperp} to sketch the range of $v_\perp$ that can be probed at a survey, given its range of operation in $\tE$ (Fig.~\ref{fig:gammaexcl_vperpvsM} left bottom panel), for a lens of a given mass.
We depict this in the right panel of Fig.~\ref{fig:gammaexcl_vperpvsM} for the surveys that yield the strongest limits -- Subaru-HSC and OGLE, as seen in the left panel -- which we will also use to show our main results.
Highlighted here are 
(i)  $v_\perp = 550$~km/s, the escape speed of the Milky Way, above which we can expect lenses to not be part of the virialized and clustered halos, 
(ii) $v_\perp = 220$~km/s, corresponding to lenses that may comprise dark matter: the range of $M$ spanned does indeed approximately correspond to the masses excluded by various surveys, as seen in, e.g., Refs.~\cite{reviewPBH:GreenKavanagh:2020jor,Green:2025dut}.
In this illustrative plot we have not accounted for the transverse motion of the lensing tube, which we will touch upon in the next subsection.

\subsection{Phase space distributions and event rates}
\label{subsec:eventrates}

For a line-of-sight density distribution of lenses $\rho_{\rm L}(\Dlens)$,  the differential rate of lenses entering a cylindrical segment of the lensing tube with transverse speed between $v_\perp$ and $v_\perp + dv_\perp$ at some lens distance between $\Dlens$ and $\Dlens + d\Dlens$ is~\cite{griest1991ApJ} 
\begin{equation}
    d\Gamma = \frac{\rho_{\rm L}(\Dlens)}{M} 2\uT\rE   v_\perp f_\perp(v_\perp) dv_\perp  d\Dlens \psi(M) dM~, 
\end{equation}
where $f_\perp$ is the distribution of transverse speeds $v_\perp$, obtained by integrating the 3D distribution $f_v(\vec{v})$ along the line of sight and multiplying the area element:
\begin{equation}
f_\perp(v_\perp) = 2\pi v_\perp \int_0^\infty dv_z f_v(v_\perp,v_z)~,
\label{eq:fperp}
\end{equation}
and
$\psi(M)$ is the lens mass distribution, which for simplicity we will take to be Dirac delta in this work.
Including the detection efficiency $\varepsilon(\tE)$ and defining $x\equiv \Dlens/\Dsource$, we now obtain
\begin{equation}
  \frac{d^2\Gamma}{dx dv_\perp} =  \varepsilon(\tE) \frac{\rho_{\rm L}(x)}{M} 2\uT\rE(x) \Dsource v_\perp(x) f_\perp (v_\perp)~. 
  \label{eq:eventratevperpgenfdistrib}
\end{equation}

Changing variables via Eq.~\eqref{eq:tEvvperp}, 
\begin{equation}
   \frac{d^2\Gamma}{dx d\tE} = \varepsilon(\tE) \frac{\rho_{\rm L}(x)}{M} \Dsource v_{\perp}^3(x) f_\perp(v_\perp)~.
   \label{eq:eventrategenfdistrib}
\end{equation}

So far in this section we have neglected transverse velocities of the source ($\vec{v}^{\rm S}_\perp$) and observer ($\vec{v}^{\rm O}_\perp$), constituting the so-called ``bulk motion'' of the lensing tube.
Including them would imply the boost $\vec{v}_\perp \to \vec{v}_\perp + x \vec{v}^{\rm S}_\perp + (1- x) \vec{v}^{\rm O}_\perp$ in the velocity distribution that gave rise to the speed distribution $f_\perp(v_\perp)$~\cite{griest1991ApJ}.
The relative transverse speed between the LMC and the Sun is about 350~km/s~\cite{griest1991ApJ}, and between M31 and the Sun is about 150~km/s~\cite{M31velocity:vanderMarel2012}.
This implies that the effect of the bulk motion on the event rate is negligible for our fast lenses with speeds $\gg 10^{-3}c$, has an $\mathcal{O}(10\%)$ effect on lenses with dark matter-like speeds of $10^{-3}c$~\cite{griest1991ApJ,Niikura:2019kqi}, and is extremely significant for lenses moving at $\ll 10^{-3}c$.
To put it differently, for slow lenses, events occur almost entirely due to the motion of the lensing tube.
The event rate, and thus the limits on lens populations, would then closely track those for $10^{-3}c$ lenses: the lensing tube bulk motion presents a microlensing ``floor'' on lens velocities.
Nevertheless, we present results in the next section pretending that $|\vec{v}_\perp + x \vec{v}^{\rm S}_\perp + (1- x) \vec{v}^{\rm O}_\perp|$ is distributed around $10^{-4}c$.
We do this mainly to illustrate the trends of our limits if the lensing tube bulk motion {\em had} been negligible, but also to cover for the (unlikely) possibility, mentioned in the Introduction, of a swarm of lenses around the lensing tube co-moving with it.

In this work we will adapt two benchmark spatial distributions $\rho_{\rm L}(x)$: one that is uniform ($\rho$ independent of $x$), and another with lenses tracking the dark matter population, which we take as the NFW galactocentric radial profile: 
\begin{eqnarray}
    \rho(r)& &= \frac{\rho_0}{(r/r_s)(1+r/r_s)^2}~,\\
\nonumber r_{\rm MW}  &=&    \sqrt{r_\odot^2 + x^2 \Dsource^2 - 2 x r_\odot \Dsource \cos\ell \cos b},\\
\nonumber r_{\rm M31}, r_{\rm LMC} &=& \Dsource (1-x)~.
    \label{eq:NFW}
\end{eqnarray}
where we take 
$\{\rho_0,r_s\}$ = 
\{0.274 ${\rm GeV/cm}^3$, 18.6 ${\rm kpc}$\} 
(\{0.189 ${\rm GeV/cm}^3$, 25 ${\rm kpc}$\})
[\{0.106 ${\rm GeV/cm}^3$, 15 ${\rm kpc}$\}] 
for the Milky Way (M31) [LMC], and the distance of the Sun from the MW centre $r_\odot = 8.2$~kpc.
We generically expect our lenses, especially those with speeds $\gtrsim  10^{-2}~c$ that exceed galactic escape speeds, to not cluster and hence be uniformly distributed in space.
Nonetheless we also show results for NFW-distributed lenses that may arise from production sites in galactic halos, e.g., mirror neutron stars that are birthed by mirror main sequence stars, and fragmented structures in dissipative dark matter models.
We also consider two qualitatively different velocity distributions $f_v(\vec{v})$, with their mean speeds varying from $10^{-4} c$ to $10^{-1} c$. 
The first is a Maxwell-Boltzmann distribution, describing an ideal virialized system; it is a Gaussian with dispersion in speeds roughly the size of the mean speed.
For our second benchmark, we pick a distribution with no dispersion in speeds, viz., a Dirac delta distribution.
While realistically we don't expect a cosmological  mechanism to produce our lenses at a single speed, this benchmark nonetheless illustrates several effects that appear as the spread/dispersion of a velocity distribution is reduced.

For the Maxwell-Boltzmann distribution, from Eqs.~\eqref{eq:fperp} and \eqref{eq:eventrategenfdistrib},
\begin{eqnarray}
 \nonumber   f^{\rm MB}_v(\vec{v}) &=& \frac{1}{(\pi v_0^2)^{3/2}}\exp(-|\vec{v}|^2/v_0^2) \\
\nonumber \Rightarrow    f^{\rm MB}_\perp(v_\perp) &=& \frac{2v_\perp}{v^2_0} \exp (-v_\perp^2/v_0^2)~\\
\Rightarrow  \frac{d^2\Gamma}{dx d\tE} &=& \varepsilon(\tE) \frac{2 \Dsource}{v_0^2M} \rho_{\rm L}(x) v_\perp^4(x) e^{-v^2_\perp(x)/v^2_0}~,
\label{eq:fvEvrateMB}
\end{eqnarray}
where the event rate obtained is as seen in the literature~\cite{CroonMcKeenRajECOlocation1,Green:2025dut}.
For the Dirac delta distribution,
\begin{eqnarray}
 \nonumber   f^{\rm DD}_v(\vec{v}) &=& \frac{1}{4\pi \bar{v}^2}\delta(|\vec{v}| - v_0) \\
\Rightarrow    f^{\rm DD}_\perp(v_\perp) &=& \nonumber \frac{v_\perp}{v_0} \frac{1}{\sqrt{v_0^2 - v_\perp^2}}~,\\
\Rightarrow  \frac{d^2\Gamma}{dx d\tE} &=& \varepsilon(\tE) \frac{\Dsource}{v_0 M} \rho_{\rm L}(x) \frac{v_\perp^4(x)}{\sqrt{v_0^2-v^2_\perp(x)}}~.
\label{eq:fvEvRateDD}
\end{eqnarray}
Note that while $v_0$ is the circular speed in $f_v^{\rm MB}$, it is the uniform speed in $f_v^{\rm DD}$.
The divergence in $f^{\rm DD}_\perp(v_\perp)$ at $v_\perp = v_0$ is an artifact of projecting a mono-energetic 3D distribution to 2D, and would be regulated in any realistic population with finite dispersion.

The number of events expected in a microlensing survey is then 
\begin{equation}
    N_{\rm ev} = N_\star T_{\rm obs} \int_0^1 dx \int_{t_{\rm E,min}}^{t_{\rm E,max}} d\tE \frac{d^2\Gamma}{dx d\tE}~,
    \label{eq:Nev}
\end{equation}
where $t_{\rm E, min}$ and $t_{\rm E, max}$ are the minimum and maximum timescales of events in the survey between which $\varepsilon(\tE) > 0$, set respectively by the observational cadence and net run-time.
We note that, when there is weak dependence on $x$ (as in the cases we consider), $N_{\rm ev}$ for a given $M$ is approximately the same for both Maxwell-Boltzmann and Dirac delta velocity distributions.
This is best seen by integrating $d^2\Gamma/dxdv_\perp$ in Eq.~\eqref{eq:eventratevperpgenfdistrib}, which gives $N_{\rm ev} \propto \int dv_\perp \varepsilon(v_\perp) f_\perp(v_\perp)$.
For $\varepsilon(v_\perp)$ that is roughly constant while it is non-zero, this becomes $N_{\rm ev} \propto \langle v_\perp \rangle$, the mean speed. 
Between Dirac delta and Maxwell-Boltzmann distributions we find that $\langle v_\perp \rangle$ differs by not more than $10-100\%$ across the surveys (showing that the shape of the distribution isn't so important as the mean), so that between the distributions $N_{\rm ev} (M)$ differs at most by a factor of 2.
(For illustration, if we set $\varepsilon = 1$ and take the integration limits as [0,$\infty$], $\langle v_\perp \rangle$ = $\sqrt{\pi}v_0/2 \simeq 0.89v_0$ for $f_v^{\rm MB}$ and $\pi v_0/4 \simeq 0.79 v_0$ for $f_v^{\rm DD}$.)

\section{Results}
\label{sec:results}

In Fig.~\ref{fig:all_limits_nfw_uniform_MB_DD} we show 95\% C.L. constraints from various surveys, setting $N_{\rm ev}$ in Eq.~\eqref{eq:Nev} to the values of $N_{\rm excl}$ in Table~\ref{tab:surveydetails} and taking the efficiencies in Fig.~\ref{fig:gammaexcl_vperpvsM}, in the plane of $\tilde{r}$ vs $M$, where $\tilde{r}$ is defined by setting $\rho_{\rm L}(x) = \tilde{r} \rho_{\rm norm}(x)$, with the normalizing density
\begin{equation}
    \rho_{\rm norm}(x) = \begin{cases}
     \rho^\odot_{\rm DM} = 0.3~{\rm GeV/cm}^3, \ {\rm flat} \ \rho(x), \\
    \rho_{\rm DM}^{\rm NFW}(x), \ {\rm NFW \ \rho(x) \ in \ Eq.~(7)}~,
    \end{cases}
    \label{eq:rhonorm}
\end{equation}
 where $\rho^\odot_{\rm DM}$ is the dark matter density at the solar vicinity.
We emphasize that $\tilde{r}$ is merely a {\em ratio} of our lens density to well-known benchmark densities that happen to be those of dark matter, and that it is \underline{not} a {\em fraction} of cold dark matter populated by our lenses: we remind the reader that our lens speeds deviate considerably from dark matter's virialized speeds. 
We choose to present our results this way to make direct visual comparisons to ``$f$ vs $M$'' style plots seen in the PBH dark matter literature.
The four panels for either choice of spatial distribution correspond to slow lenses ($v_0 = 10^{-4}c$), dark matter-speed lenses ($v_0 = 7.3 \times 10^{-4}c = 220$~km/s), 
and fast lenses ($v_0 = 10^{-2}c,10^{-1}c$).
We emphasize again that the limits for the slow lenses are likely unphysical, as the $\sim10^{-3}c$ speed transverse motion of the lensing tube would determine their event rate.
We show them here primarily to illustrate the trends of our limits, but also to cover for the scenarios of non-isotropic lenses mentioned in the Introduction and Sec.~\ref{subsec:eventrates}.
The limits for $v_0 = 220$~km/s with Maxwell-Boltzmann velocities and NFW densities closely agree with those on point-like dark matter lenses seen in the literature~\cite{review:Green:2026xhw}.
For Subaru-HSC and OGLE-hc we take into account the effects of the finite extent of the source by using the magnification derived in Ref.~\cite{wittmao1994} and solving for $\uT$\footnote{Amusingly, we obtained a close reproduction of our results when we set $\uT$ to just $\min[1,\theta_\star/\thetaE]$, where $\theta_\star$ is the angular extent of the source star and $\thetaE$ the Einstein angle $\rE/\Dlens$.}, and by setting the radii of source stars to 1~$R_\odot$ which closely reproduces limits that account for the stellar radius distribution~\cite{SubaruHSC:Niikura:2017zjd,Smyth:2019whb,CroonMcKeenRajECOlocation2,OGLELMChc:Mroz:2024wia,reviewPBH:GreenKavanagh:2020jor}. 
(The wave optics effect, operative at Subaru-HSC, is sub-dominant to the finite source effect~\cite{SubaruHSC:Niikura:2017zjd,CroonMcKeenRajECOlocation2}, so we neglect it here.)
The wiggles seen in the Subaru-HSC limits are numerical artifacts of incorporating this effect.
Our use of efficiency curves corresponding to source stars of a single luminosity/size introduces an $\mathcal{O}(10\%)$ error in our limits that is not visible in our log-log plots.
The Subaru-HSC efficiencies are derived as a function of the full width at half-maximum of the light-curve, given by $t_{\rm FWHM} \simeq \sqrt{3} u \tE$ for impact parameters (in $\rE$ units) $u\ll1$~\cite{BaltzSilk:1999fr}; as the $u$ distribution is unknown, we simply take $t_{\rm FWHM} = \tE$.
While we have used the efficiencies and Subaru-HSC dataset of Ref.~\cite{SubaruHSC:Niikura:2017zjd} to show our limits, a currently debated re-analysis of the Subaru-HSC pipeline with an updated dataset has appeared~\cite{SubaruHSC:Sugiyama:2026kpv}.
We discuss this in more detail in Sec.~\ref{sec:discs}.

We notice that the limits for Dirac delta- and Maxwell-Boltzmann-distributed lenses are close to each other, which is due to the approximate equality in event count discussed at the end of the previous section.
The Dirac delta limits are slightly stronger at the deepest point and slightly weaker elsewhere; this is due to the Maxwell-Boltzmann event rates getting contributions from the tails of the 2D distribution $f_\perp$, which is not as pronounced for the Dirac delta case.
As we scan down the panels, we also see the limits shifting to higher $M$, which is due to the $v_\perp$-$M$ degeneracy at fixed $\tE$ seen in Fig.~\ref{fig:gammaexcl_vperpvsM} right panel; indeed, the ranges of $M$ spanned at fixed $v_\perp$ in that figure correspond roughly to the ranges of $M$ spanned in Fig.~\ref{fig:all_limits_nfw_uniform_MB_DD}. 
Moreover, we see that the limits from a given survey on $M$ scale as $M \propto v_0^2$, which follows from $N_{\rm ev} \propto \langle v_\perp \rangle/\sqrt{M}$ as obtained from Eq.~\eqref{eq:eventratevperpgenfdistrib}, using the definition of $\rE$ and integrating over $v_\perp$.
The Subaru-HSC limits show interesting behavior in both plot columns.
While the strongest limits on $\tilde{r}$ from OGLE-hc and OGLE remain the same as $v_0$ increases, those on Subaru-HSC deepen, and in the case of NFW lenses, even outdo the other surveys.
This is because the finite source effect at Subaru-HSC -- which serves to suppress microlensing magnifications and hence event rates -- weakens as the range of $M$ constrained increases: the concomitant increase in $\rE$ makes the Einstein angle larger than the angular size of source stars.
The finite source effect is also in play at OGLE-hc but at higher $M$ than for Subaru-HSC since the microlensing baseline is smaller. 
The effect can also be seen in the left edge of the Subaru-HSC and OGLE-hc limits: steeper in the upper panels, and becoming more parallel to the left edge of the OGLE limits in the lower panels.
For the NFW-like distribution the deepest point of all three limits, and for the uniform distribution the two deepest OGLE limits, are all within an order of magnitude, which derives from the scaling $N_{\rm ev} \propto T_{\rm obs}/\sqrt{M}$.
However, for the uniform distribution the strongest limits on $\tilde{r}$ from Subaru-HSC are 1$-$3 orders deeper than the strongest OGLE limits.
This is due to a combination of the above effect of source-size finiteness as well as the fact that, for flat spatial distributions, the optical depth of lenses toward M31 is much greater than toward the LMC.

The above discussion highlights an important effect and a key finding of our study.
The finite source effect (and in the case of Subaru-HSC, wave optics effect as well) is an unfortunate ``wall" for microlensing searches of dark matter in the form of PBHs and other macroscopic objects, and defines the right edge of the so-called PBH mass window. 
Increasing the observing cadence at surveys like Subaru-HSC and OGLE would not necessarily probe lens masses below the values at which this effect kicks in.
However, there is no such drawback when microlensing measurements are interpreted in terms of our fast ($\gg 10^{-3}$c) lenses: efforts to make observations at higher cadence {\em do} push the low mass end.
This should incentivize microlensing collaborations to carry out surveys at the highest cadence possible.

\section{Discussion}
\label{sec:discs}

In this study we set limits on the populations of exotic compact objects, not necessarily {\em the} dark matter, with average speeds ranging from $10^{-4}c-10^{-1}c$, under some assumptions on their phase space distribution using observations at the microlensing surveys Subaru-HSC and OGLE.
While this work was in progress, an update to the 7 hr-runtime Subaru-HSC limits of Ref.~\cite{SubaruHSC:Niikura:2017zjd}, appeared in Ref.~\cite{SubaruHSC:Sugiyama:2026kpv} with a total run-time of 39.3 hours. 
Despite the increased exposure, the upper limits on PBH populations {\em decreased} considerably, with the minimum constrained mass shifting to $\sim10^{-10}~M_\odot$, thus seemingly widening the PBH dark matter mass window by an order of magnitude. 
This was chiefly due to incorporation of finite source effects into simulations to determine the detection efficiencies and stricter selection cuts for microlensing light-curves, all of which resulted in the efficiencies dropping by a factor $>6$ in comparison to Ref.~\cite{SubaruHSC:Niikura:2017zjd}.
This analysis identified 12 microlensing events in their dataset, which when interpreted as positives yielded best-fit PBH masses of $10^{-8}M_\odot-10^{-6}M_\odot$.
This was followed by another inspection of the updated Subaru-HSC dataset by Ref.~\cite{SubaruHSC:Mroz:2026nez}, which found that all the 12 abovementioned events can be attributed to variable stars.
Due to the high number of false positives, this study called into question the accuracy of the selection criteria of Ref.~\cite{SubaruHSC:Sugiyama:2026kpv} that determined the efficiencies, and hence the robustness of the limits.
Further, the authors of Ref.~\cite{SubaruHSC:Mroz:2026nez} mention plans to re-analyze the entire Subaru-HSC dataset.
As the Subaru-HSC results depend on an ongoing debate, in our work we have taken a pragmatic approach,  estimating our limits using efficiencies and exposures of the old dataset in Ref.~\cite{SubaruHSC:Niikura:2017zjd}, and will update our results when the microlensing observers community reaches a consensus on the Subaru-HSC pipeline.

As we had mentioned in the Introduction, the mass-speed degeneracy of photometric microlensing may be broken with astrometric microlensing.
It may also be broken by determining the lens mass when finite source effects are in play~\cite{2017Univ....3...53L}.
It can be shown from the differential rate in Ref.~\cite{griest1991ApJ} that, for impact parameters $u$~\cite{VermaTamtaRajhaloindep},
\begin{eqnarray}
 \frac{d\Gamma}{du} &=& \sqrt{\frac{4G\Dsource^3}{c^2M}} \tilde\mu~,\\   
\nonumber \tilde{\mu} &=& \int_0^1 dx \int dv_\perp \sqrt{x(1-x)} \rho(x) v_\perp^2(x) f_\perp(v_\perp(x))~.
\end{eqnarray} 
Thus for a given $M$, observed event distributions of the impact parameter (measured from light curves) would give us the quantity $\tilde\mu$ that integrates over (the unknown) phase space distributions of the lenses.
If the lenses are known to be uniformly distributed in space and $v_\perp$ is independent of $x$, $\rho(x)$ can be factored out and the squared mean speed of the lenses can be obtained.
Further, if futuristic proposals using ``femtolensing'' and ``picolensing'' techniques with gamma-ray bursts as sources at cosmological baselines~\cite{femtolensing:Katz:2018JCAP,femtolensing:Jung:2020PRR,picolensingDaksha:Gawade:2023gmt,picolensing:Fedderke:2025PRD} come to fruition, the masses of the lenses can be inferred regardless of their velocity.
That is, while the photometric microlensing event rate goes as $\sim \rho_{\rm L}\langle v_\perp\rangle$, the event rate in these techniques $\propto \rho_{\rm L}$.
Here we draw attention to the estimate of the population of asteroids in Ref.~\cite{picolensing:Fedderke:2025PRD}, with the finding that they make up no more than $10^{-7}$ times the total stellar mass of galaxies.
This is below microlensing sensitivities for any velocity, however it is interesting to note that our results imply that a complex dark sector producing free-floating asteroid-mass objects at high velocities may be within the reach of microlensing surveys.

In that vein, we note that fast and slow lenses may turn up at current and near-future missions capable of microlensing.
Sensitivities to $10^{-3}c$ speed dark compact objects have been derived for Rubin~\cite{darkdisk:WinchSetfordCurtin:2020cju,RubinreachDrlicaWagner:2019mwo,RubinreachCroonRomao:2025kxx},
Roman~\cite{Romanreach:DeRoccoProfumoSmyth:2023gde} (also in astrometric microlensing~\cite{Romanreachastrometric:fardeen2024}),
Euclid~\cite{Euclidreach:Hamolli2021},
a Euclid-Roman joint survey~\cite{EuclidRomanjointreach:Bachelet2022}, 
and current and future x-ray telescopes~\cite{xraymicrolensing:Bai:2018bej,xraymicrolensing:Tamta:2024pow}.
These reach new parameter space, which we expect for our lenses too.
Our treatment can be extended to structures with spatial extent comparable to the Einstein radius, which modifies the light curve compared to point-like lenses and the threshold impact parameter $\uT$, hence the event rate and limits~\cite{CroonMcKeenRajECOlocation1,CroonMcKeenRajECOlocation2}. 
Our constraints from OGLE reach lens masses $\gg M_\odot$, where we expect complementary constraints from such dynamical effects as heating of stellar populations in ultra-faint dwarf galaxies~\cite{Graham:2023unf} and
disruption of wide binaries~\cite{Yoo:2003fr}, and
the accretion of baryonic matter on to the lenses leading to heating of gas in dwarf galaxies~\cite{Lu:2020bmd}, 
x-ray and radio emission~\cite{Inoue:2017csr,Manshanden:2018tze}, and
effects on the ionization history of the universe leaving an imprint on the CMB~\cite{Croon:2024rmw}; see Ref.~\cite{reviewPBH:GreenKavanagh:2020jor} for other pertinent references.
To our knowledge, these have not been worked out for super-massive objects with non-dark matter phase space distributions.
We leave to future work these and the other investigations discussed above.

\section*{Acknowledgments}
N. R. acknowledges support from the grant ANRF/ECRG/2024/000387/PMS and the Infosys Foundation, Bangalore.
H. V. is supported by the funding for the Roman Galactic Exoplanet Survey Project Infrastructure Team (NASA grant 80NSSC24M0022).

\bibliography{refs-nonthermlens}

\begin{thebibliography}{71}%
\makeatletter
\providecommand \@ifxundefined [1]{%
 \@ifx{#1\undefined}
}%
\providecommand \@ifnum [1]{%
 \ifnum #1\expandafter \@firstoftwo
 \else \expandafter \@secondoftwo
 \fi
}%
\providecommand \@ifx [1]{%
 \ifx #1\expandafter \@firstoftwo
 \else \expandafter \@secondoftwo
 \fi
}%
\providecommand \natexlab [1]{#1}%
\providecommand \enquote  [1]{``#1''}%
\providecommand \bibnamefont  [1]{#1}%
\providecommand \bibfnamefont [1]{#1}%
\providecommand \citenamefont [1]{#1}%
\providecommand \href@noop [0]{\@secondoftwo}%
\providecommand \href [0]{\begingroup \@sanitize@url \@href}%
\providecommand \@href[1]{\@@startlink{#1}\@@href}%
\providecommand \@@href[1]{\endgroup#1\@@endlink}%
\providecommand \@sanitize@url [0]{\catcode `\\12\catcode `\$12\catcode `\&12\catcode `\#12\catcode `\^12\catcode `\_12\catcode `\%12\relax}%
\providecommand \@@startlink[1]{}%
\providecommand \@@endlink[0]{}%
\providecommand \url  [0]{\begingroup\@sanitize@url \@url }%
\providecommand \@url [1]{\endgroup\@href {#1}{\urlprefix }}%
\providecommand \urlprefix  [0]{URL }%
\providecommand \Eprint [0]{\href }%
\providecommand \doibase [0]{http://dx.doi.org/}%
\providecommand \selectlanguage [0]{\@gobble}%
\providecommand \bibinfo  [0]{\@secondoftwo}%
\providecommand \bibfield  [0]{\@secondoftwo}%
\providecommand \translation [1]{[#1]}%
\providecommand \BibitemOpen [0]{}%
\providecommand \bibitemStop [0]{}%
\providecommand \bibitemNoStop [0]{.\EOS\space}%
\providecommand \EOS [0]{\spacefactor3000\relax}%
\providecommand \BibitemShut  [1]{\csname bibitem#1\endcsname}%
\let\auto@bib@innerbib\@empty
\bibitem [{\citenamefont {Einstein}(1936)}]{Einstein:1936llh}%
  \BibitemOpen
  \bibfield  {author} {\bibinfo {author} {\bibfnamefont {A.}~\bibnamefont {Einstein}},\ }\href {\doibase 10.1126/science.84.2188.506} {\bibfield  {journal} {\bibinfo  {journal} {Science}\ }\textbf {\bibinfo {volume} {84}},\ \bibinfo {pages} {506} (\bibinfo {year} {1936})}\BibitemShut {NoStop}%
\bibitem [{\citenamefont {Narayan}\ and\ \citenamefont {Bartelmann}(1996)}]{NarayanBartelmann}%
  \BibitemOpen
  \bibfield  {author} {\bibinfo {author} {\bibfnamefont {R.}~\bibnamefont {Narayan}}\ and\ \bibinfo {author} {\bibfnamefont {M.}~\bibnamefont {Bartelmann}},\ }in\ \href@noop {} {\emph {\bibinfo {booktitle} {{13th Jerusalem Winter School in Theoretical Physics: Formation of Structure in the Universe}}}}\ (\bibinfo {year} {1996})\ \Eprint {http://arxiv.org/abs/astro-ph/9606001} {arXiv:astro-ph/9606001} \BibitemShut {NoStop}%
\bibitem [{\citenamefont {Green}\ and\ \citenamefont {Kavanagh}(2021)}]{reviewPBH:GreenKavanagh:2020jor}%
  \BibitemOpen
  \bibfield  {author} {\bibinfo {author} {\bibfnamefont {A.~M.}\ \bibnamefont {Green}}\ and\ \bibinfo {author} {\bibfnamefont {B.~J.}\ \bibnamefont {Kavanagh}},\ }\href {\doibase 10.1088/1361-6471/abc534} {\bibfield  {journal} {\bibinfo  {journal} {J. Phys. G}\ }\textbf {\bibinfo {volume} {48}},\ \bibinfo {pages} {043001} (\bibinfo {year} {2021})},\ \Eprint {http://arxiv.org/abs/2007.10722} {arXiv:2007.10722 [astro-ph.CO]} \BibitemShut {NoStop}%
\bibitem [{\citenamefont {Green}(2025)}]{Green:2025dut}%
  \BibitemOpen
  \bibfield  {author} {\bibinfo {author} {\bibfnamefont {A.~M.}\ \bibnamefont {Green}},\ }\href {\doibase 10.1088/1475-7516/2025/04/023} {\bibfield  {journal} {\bibinfo  {journal} {JCAP}\ }\textbf {\bibinfo {volume} {04}},\ \bibinfo {pages} {023} (\bibinfo {year} {2025})},\ \Eprint {http://arxiv.org/abs/2501.02610} {arXiv:2501.02610 [astro-ph.GA]} \BibitemShut {NoStop}%
\bibitem [{\citenamefont {Henriksen}\ and\ \citenamefont {Widrow}(1995)}]{WidrowHClouds}%
  \BibitemOpen
  \bibfield  {author} {\bibinfo {author} {\bibfnamefont {R.~N.}\ \bibnamefont {Henriksen}}\ and\ \bibinfo {author} {\bibfnamefont {L.~M.}\ \bibnamefont {Widrow}},\ }\href {\doibase 10.1086/175336} {\bibfield  {journal} {\bibinfo  {journal} {Astrophys. J.}\ }\textbf {\bibinfo {volume} {441}},\ \bibinfo {pages} {70} (\bibinfo {year} {1995})},\ \Eprint {http://arxiv.org/abs/astro-ph/9402002} {arXiv:astro-ph/9402002 [astro-ph]} \BibitemShut {NoStop}%
\bibitem [{\citenamefont {Fairbairn}\ \emph {et~al.}(2017)\citenamefont {Fairbairn}, \citenamefont {Marsh},\ and\ \citenamefont {Quevillon}}]{Fairbairn:2017dmf}%
  \BibitemOpen
  \bibfield  {author} {\bibinfo {author} {\bibfnamefont {M.}~\bibnamefont {Fairbairn}}, \bibinfo {author} {\bibfnamefont {D.~J.~E.}\ \bibnamefont {Marsh}}, \ and\ \bibinfo {author} {\bibfnamefont {J.}~\bibnamefont {Quevillon}},\ }\href {\doibase 10.1103/PhysRevLett.119.021101} {\bibfield  {journal} {\bibinfo  {journal} {Phys. Rev. Lett.}\ }\textbf {\bibinfo {volume} {119}},\ \bibinfo {pages} {021101} (\bibinfo {year} {2017})},\ \Eprint {http://arxiv.org/abs/1701.04787} {arXiv:1701.04787 [astro-ph.CO]} \BibitemShut {NoStop}%
\bibitem [{\citenamefont {Blinov}\ \emph {et~al.}(2020)\citenamefont {Blinov}, \citenamefont {Dolan},\ and\ \citenamefont {Draper}}]{Blinov_2020}%
  \BibitemOpen
  \bibfield  {author} {\bibinfo {author} {\bibfnamefont {N.}~\bibnamefont {Blinov}}, \bibinfo {author} {\bibfnamefont {M.~J.}\ \bibnamefont {Dolan}}, \ and\ \bibinfo {author} {\bibfnamefont {P.}~\bibnamefont {Draper}},\ }\href {\doibase 10.1103/physrevd.101.035002} {\bibfield  {journal} {\bibinfo  {journal} {Physical Review D}\ }\textbf {\bibinfo {volume} {101}} (\bibinfo {year} {2020}),\ 10.1103/physrevd.101.035002}\BibitemShut {NoStop}%
\bibitem [{\citenamefont {Croon}\ \emph {et~al.}(2020{\natexlab{a}})\citenamefont {Croon}, \citenamefont {McKeen}, \citenamefont {Raj},\ and\ \citenamefont {Wang}}]{CroonMcKeenRajECOlocation2}%
  \BibitemOpen
  \bibfield  {author} {\bibinfo {author} {\bibfnamefont {D.}~\bibnamefont {Croon}}, \bibinfo {author} {\bibfnamefont {D.}~\bibnamefont {McKeen}}, \bibinfo {author} {\bibfnamefont {N.}~\bibnamefont {Raj}}, \ and\ \bibinfo {author} {\bibfnamefont {Z.}~\bibnamefont {Wang}},\ }\href {\doibase 10.1103/PhysRevD.102.083021} {\bibfield  {journal} {\bibinfo  {journal} {Phys. Rev. D}\ }\textbf {\bibinfo {volume} {102}},\ \bibinfo {pages} {083021} (\bibinfo {year} {2020}{\natexlab{a}})},\ \Eprint {http://arxiv.org/abs/2007.12697} {arXiv:2007.12697 [astro-ph.CO]} \BibitemShut {NoStop}%
\bibitem [{\citenamefont {Croon}\ \emph {et~al.}(2020{\natexlab{b}})\citenamefont {Croon}, \citenamefont {McKeen},\ and\ \citenamefont {Raj}}]{CroonMcKeenRajECOlocation1}%
  \BibitemOpen
  \bibfield  {author} {\bibinfo {author} {\bibfnamefont {D.}~\bibnamefont {Croon}}, \bibinfo {author} {\bibfnamefont {D.}~\bibnamefont {McKeen}}, \ and\ \bibinfo {author} {\bibfnamefont {N.}~\bibnamefont {Raj}},\ }\href {\doibase 10.1103/PhysRevD.101.083013} {\bibfield  {journal} {\bibinfo  {journal} {Phys. Rev. D}\ }\textbf {\bibinfo {volume} {101}},\ \bibinfo {pages} {083013} (\bibinfo {year} {2020}{\natexlab{b}})}\BibitemShut {NoStop}%
\bibitem [{\citenamefont {Bai}\ \emph {et~al.}(2020)\citenamefont {Bai}, \citenamefont {Long},\ and\ \citenamefont {Lu}}]{BaiDMACHOs:2020jfm}%
  \BibitemOpen
  \bibfield  {author} {\bibinfo {author} {\bibfnamefont {Y.}~\bibnamefont {Bai}}, \bibinfo {author} {\bibfnamefont {A.~J.}\ \bibnamefont {Long}}, \ and\ \bibinfo {author} {\bibfnamefont {S.}~\bibnamefont {Lu}},\ }\href {\doibase 10.1088/1475-7516/2020/09/044} {\bibfield  {journal} {\bibinfo  {journal} {JCAP}\ }\textbf {\bibinfo {volume} {09}},\ \bibinfo {pages} {044} (\bibinfo {year} {2020})},\ \Eprint {http://arxiv.org/abs/2003.13182} {arXiv:2003.13182 [astro-ph.CO]} \BibitemShut {NoStop}%
\bibitem [{\citenamefont {Ansari}\ \emph {et~al.}(2024)\citenamefont {Ansari}, \citenamefont {Singh~Bhandari},\ and\ \citenamefont {Thalapillil}}]{AnsariArunQBalls:2023cay}%
  \BibitemOpen
  \bibfield  {author} {\bibinfo {author} {\bibfnamefont {A.}~\bibnamefont {Ansari}}, \bibinfo {author} {\bibfnamefont {L.}~\bibnamefont {Singh~Bhandari}}, \ and\ \bibinfo {author} {\bibfnamefont {A.~M.}\ \bibnamefont {Thalapillil}},\ }\href {\doibase 10.1103/PhysRevD.109.023003} {\bibfield  {journal} {\bibinfo  {journal} {Phys. Rev. D}\ }\textbf {\bibinfo {volume} {109}},\ \bibinfo {pages} {023003} (\bibinfo {year} {2024})},\ \Eprint {http://arxiv.org/abs/2302.11590} {arXiv:2302.11590 [hep-ph]} \BibitemShut {NoStop}%
\bibitem [{\citenamefont {Sajadian}(2021)}]{FFPRoman:Sajadian2021}%
  \BibitemOpen
  \bibfield  {author} {\bibinfo {author} {\bibfnamefont {S.}~\bibnamefont {Sajadian}},\ }\href {\doibase 10.1093/mnras/stab1907} {\bibfield  {journal} {\bibinfo  {journal} {Monthly Notices of the Royal Astronomical Society}\ }\textbf {\bibinfo {volume} {506}},\ \bibinfo {pages} {3615–3628} (\bibinfo {year} {2021})}\BibitemShut {NoStop}%
\bibitem [{\citenamefont {Bachelet}\ \emph {et~al.}(2022)\citenamefont {Bachelet}, \citenamefont {Specht}, \citenamefont {Penny}, \citenamefont {Hundertmark}, \citenamefont {Awiphan}, \citenamefont {Beaulieu}, \citenamefont {Dominik}, \citenamefont {Kerins}, \citenamefont {Maoz}, \citenamefont {Meade}, \citenamefont {Nucita}, \citenamefont {Poleski}, \citenamefont {Ranc}, \citenamefont {Rhodes},\ and\ \citenamefont {Robin}}]{EuclidRomanjointreach:Bachelet2022}%
  \BibitemOpen
  \bibfield  {author} {\bibinfo {author} {\bibfnamefont {E.}~\bibnamefont {Bachelet}}, \bibinfo {author} {\bibfnamefont {D.}~\bibnamefont {Specht}}, \bibinfo {author} {\bibfnamefont {M.}~\bibnamefont {Penny}}, \bibinfo {author} {\bibfnamefont {M.}~\bibnamefont {Hundertmark}}, \bibinfo {author} {\bibfnamefont {S.}~\bibnamefont {Awiphan}}, \bibinfo {author} {\bibfnamefont {J.-P.}\ \bibnamefont {Beaulieu}}, \bibinfo {author} {\bibfnamefont {M.}~\bibnamefont {Dominik}}, \bibinfo {author} {\bibfnamefont {E.}~\bibnamefont {Kerins}}, \bibinfo {author} {\bibfnamefont {D.}~\bibnamefont {Maoz}}, \bibinfo {author} {\bibfnamefont {E.}~\bibnamefont {Meade}}, \bibinfo {author} {\bibfnamefont {A.~A.}\ \bibnamefont {Nucita}}, \bibinfo {author} {\bibfnamefont {R.}~\bibnamefont {Poleski}}, \bibinfo {author} {\bibfnamefont {C.}~\bibnamefont {Ranc}}, \bibinfo {author} {\bibfnamefont {J.}~\bibnamefont {Rhodes}}, \ and\ \bibinfo {author} {\bibfnamefont {A.~C.}\ \bibnamefont {Robin}},\ }\href {\doibase 10.1051/0004-6361/202140351}
  {\bibfield  {journal} {\bibinfo  {journal} {Astronomy \& Astrophysics}\ }\textbf {\bibinfo {volume} {664}},\ \bibinfo {pages} {A136} (\bibinfo {year} {2022})}\BibitemShut {NoStop}%
\bibitem [{\citenamefont {Niikura}\ \emph {et~al.}(2019{\natexlab{a}})\citenamefont {Niikura} \emph {et~al.}}]{SubaruHSC:Niikura:2017zjd}%
  \BibitemOpen
  \bibfield  {author} {\bibinfo {author} {\bibfnamefont {H.}~\bibnamefont {Niikura}} \emph {et~al.},\ }\href {\doibase 10.1038/s41550-019-0723-1} {\bibfield  {journal} {\bibinfo  {journal} {Nature Astron.}\ }\textbf {\bibinfo {volume} {3}},\ \bibinfo {pages} {524} (\bibinfo {year} {2019}{\natexlab{a}})},\ \Eprint {http://arxiv.org/abs/1701.02151} {arXiv:1701.02151 [astro-ph.CO]} \BibitemShut {NoStop}%
\bibitem [{\citenamefont {Mr\'oz}\ \emph {et~al.}(2024{\natexlab{a}})\citenamefont {Mr\'oz} \emph {et~al.}}]{OGLELMChc:Mroz:2024wia}%
  \BibitemOpen
  \bibfield  {author} {\bibinfo {author} {\bibfnamefont {P.}~\bibnamefont {Mr\'oz}} \emph {et~al.},\ }\href {\doibase 10.3847/2041-8213/ad8e68} {\bibfield  {journal} {\bibinfo  {journal} {Astrophys. J. Lett.}\ }\textbf {\bibinfo {volume} {976}},\ \bibinfo {pages} {L19} (\bibinfo {year} {2024}{\natexlab{a}})},\ \Eprint {http://arxiv.org/abs/2410.06251} {arXiv:2410.06251 [astro-ph.CO]} \BibitemShut {NoStop}%
\bibitem [{\citenamefont {Mr\'oz}\ \emph {et~al.}(2024{\natexlab{b}})\citenamefont {Mr\'oz} \emph {et~al.}}]{OGLELMClc:Mroz:2024wag}%
  \BibitemOpen
  \bibfield  {author} {\bibinfo {author} {\bibfnamefont {P.}~\bibnamefont {Mr\'oz}} \emph {et~al.},\ }\href {\doibase 10.3847/1538-4365/ad452e} {\bibfield  {journal} {\bibinfo  {journal} {Astrophys. J. Suppl.}\ }\textbf {\bibinfo {volume} {273}},\ \bibinfo {pages} {4} (\bibinfo {year} {2024}{\natexlab{b}})},\ \Eprint {http://arxiv.org/abs/2403.02398} {arXiv:2403.02398 [astro-ph.GA]} \BibitemShut {NoStop}%
\bibitem [{\citenamefont {Tisserand}\ \emph {et~al.}(2007)\citenamefont {Tisserand} \emph {et~al.}}]{EROS2Tisserand}%
  \BibitemOpen
  \bibfield  {author} {\bibinfo {author} {\bibfnamefont {P.}~\bibnamefont {Tisserand}} \emph {et~al.} (\bibinfo {collaboration} {EROS-2}),\ }\href {\doibase 10.1051/0004-6361:20066017} {\bibfield  {journal} {\bibinfo  {journal} {Astron. Astrophys.}\ }\textbf {\bibinfo {volume} {469}},\ \bibinfo {pages} {387} (\bibinfo {year} {2007})},\ \Eprint {http://arxiv.org/abs/astro-ph/0607207} {arXiv:astro-ph/0607207 [astro-ph]} \BibitemShut {NoStop}%
\bibitem [{\citenamefont {Alcock}\ \emph {et~al.}(2000)\citenamefont {Alcock} \emph {et~al.}}]{MACHO:2000qbb}%
  \BibitemOpen
  \bibfield  {author} {\bibinfo {author} {\bibfnamefont {C.}~\bibnamefont {Alcock}} \emph {et~al.} (\bibinfo {collaboration} {MACHO}),\ }\href {\doibase 10.1086/309512} {\bibfield  {journal} {\bibinfo  {journal} {Astrophys. J.}\ }\textbf {\bibinfo {volume} {542}},\ \bibinfo {pages} {281} (\bibinfo {year} {2000})},\ \Eprint {http://arxiv.org/abs/astro-ph/0001272} {arXiv:astro-ph/0001272} \BibitemShut {NoStop}%
\bibitem [{\citenamefont {Jenkins}\ and\ \citenamefont {Sakellariadou}(2020)}]{fastlenses:PBHcosmicstring:Jenkins:2020ctp}%
  \BibitemOpen
  \bibfield  {author} {\bibinfo {author} {\bibfnamefont {A.~C.}\ \bibnamefont {Jenkins}}\ and\ \bibinfo {author} {\bibfnamefont {M.}~\bibnamefont {Sakellariadou}},\ }\href@noop {} {\  (\bibinfo {year} {2020})},\ \Eprint {http://arxiv.org/abs/2006.16249} {arXiv:2006.16249 [astro-ph.CO]} \BibitemShut {NoStop}%
\bibitem [{\citenamefont {Hippert}\ \emph {et~al.}(2022)\citenamefont {Hippert}, \citenamefont {Setford}, \citenamefont {Tan}, \citenamefont {Curtin}, \citenamefont {Noronha-Hostler},\ and\ \citenamefont {Yunes}}]{fastlenses:mirrorNS:Hippert:2021fch}%
  \BibitemOpen
  \bibfield  {author} {\bibinfo {author} {\bibfnamefont {M.}~\bibnamefont {Hippert}}, \bibinfo {author} {\bibfnamefont {J.}~\bibnamefont {Setford}}, \bibinfo {author} {\bibfnamefont {H.}~\bibnamefont {Tan}}, \bibinfo {author} {\bibfnamefont {D.}~\bibnamefont {Curtin}}, \bibinfo {author} {\bibfnamefont {J.}~\bibnamefont {Noronha-Hostler}}, \ and\ \bibinfo {author} {\bibfnamefont {N.}~\bibnamefont {Yunes}},\ }\href {\doibase 10.1103/PhysRevD.106.035025} {\bibfield  {journal} {\bibinfo  {journal} {Phys. Rev. D}\ }\textbf {\bibinfo {volume} {106}},\ \bibinfo {pages} {035025} (\bibinfo {year} {2022})},\ \Eprint {http://arxiv.org/abs/2103.01965} {arXiv:2103.01965 [astro-ph.HE]} \BibitemShut {NoStop}%
\bibitem [{\citenamefont {Disberg}\ and\ \citenamefont {Mandel}(2025)}]{Disberg:2025xoh}%
  \BibitemOpen
  \bibfield  {author} {\bibinfo {author} {\bibfnamefont {P.}~\bibnamefont {Disberg}}\ and\ \bibinfo {author} {\bibfnamefont {I.}~\bibnamefont {Mandel}},\ }\href {\doibase 10.3847/2041-8213/adf286} {\bibfield  {journal} {\bibinfo  {journal} {Astrophys. J. Lett.}\ }\textbf {\bibinfo {volume} {989}},\ \bibinfo {pages} {L8} (\bibinfo {year} {2025})},\ \Eprint {http://arxiv.org/abs/2505.22102} {arXiv:2505.22102 [astro-ph.HE]} \BibitemShut {NoStop}%
\bibitem [{\citenamefont {Gerosa}\ and\ \citenamefont {Moore}(2016)}]{fastlenses:BHGWsuperkix:Gerosa:2016vip}%
  \BibitemOpen
  \bibfield  {author} {\bibinfo {author} {\bibfnamefont {D.}~\bibnamefont {Gerosa}}\ and\ \bibinfo {author} {\bibfnamefont {C.~J.}\ \bibnamefont {Moore}},\ }\href {\doibase 10.1103/PhysRevLett.117.011101} {\bibfield  {journal} {\bibinfo  {journal} {Phys. Rev. Lett.}\ }\textbf {\bibinfo {volume} {117}},\ \bibinfo {pages} {011101} (\bibinfo {year} {2016})},\ \Eprint {http://arxiv.org/abs/1606.04226} {arXiv:1606.04226 [gr-qc]} \BibitemShut {NoStop}%
\bibitem [{\citenamefont {Bamber}\ \emph {et~al.}(2025)\citenamefont {Bamber}, \citenamefont {Shapiro}, \citenamefont {Ruiz},\ and\ \citenamefont {Tsokaros}}]{fastlenses:BHslingshot:Bamber:2025gxj}%
  \BibitemOpen
  \bibfield  {author} {\bibinfo {author} {\bibfnamefont {J.}~\bibnamefont {Bamber}}, \bibinfo {author} {\bibfnamefont {S.~L.}\ \bibnamefont {Shapiro}}, \bibinfo {author} {\bibfnamefont {M.}~\bibnamefont {Ruiz}}, \ and\ \bibinfo {author} {\bibfnamefont {A.}~\bibnamefont {Tsokaros}},\ }\href {\doibase 10.1103/8z1j-rzmx} {\bibfield  {journal} {\bibinfo  {journal} {Phys. Rev. D}\ }\textbf {\bibinfo {volume} {112}},\ \bibinfo {pages} {024046} (\bibinfo {year} {2025})},\ \Eprint {http://arxiv.org/abs/2505.01495} {arXiv:2505.01495 [gr-qc]} \BibitemShut {NoStop}%
\bibitem [{\citenamefont {Gresham}\ \emph {et~al.}(2023)\citenamefont {Gresham}, \citenamefont {Lee},\ and\ \citenamefont {Zurek}}]{fastlenses:longrangenuggets:Gresham:2022biw}%
  \BibitemOpen
  \bibfield  {author} {\bibinfo {author} {\bibfnamefont {M.~I.}\ \bibnamefont {Gresham}}, \bibinfo {author} {\bibfnamefont {V.~S.~H.}\ \bibnamefont {Lee}}, \ and\ \bibinfo {author} {\bibfnamefont {K.~M.}\ \bibnamefont {Zurek}},\ }\href {\doibase 10.1088/1475-7516/2023/02/048} {\bibfield  {journal} {\bibinfo  {journal} {JCAP}\ }\textbf {\bibinfo {volume} {02}},\ \bibinfo {pages} {048} (\bibinfo {year} {2023})},\ \Eprint {http://arxiv.org/abs/2209.03963} {arXiv:2209.03963 [astro-ph.HE]} \BibitemShut {NoStop}%
\bibitem [{\citenamefont {van Dokkum}\ \emph {et~al.}(2026)\citenamefont {van Dokkum}, \citenamefont {Jennings}, \citenamefont {Pasha}, \citenamefont {Conroy}, \citenamefont {Kaul}, \citenamefont {Abraham}, \citenamefont {Danieli}, \citenamefont {Romanowsky},\ and\ \citenamefont {Tremblay}}]{fastlenses:GWrecoilJWST:2025bah}%
  \BibitemOpen
  \bibfield  {author} {\bibinfo {author} {\bibfnamefont {P.}~\bibnamefont {van Dokkum}}, \bibinfo {author} {\bibfnamefont {C.}~\bibnamefont {Jennings}}, \bibinfo {author} {\bibfnamefont {I.}~\bibnamefont {Pasha}}, \bibinfo {author} {\bibfnamefont {C.}~\bibnamefont {Conroy}}, \bibinfo {author} {\bibfnamefont {I.}~\bibnamefont {Kaul}}, \bibinfo {author} {\bibfnamefont {R.}~\bibnamefont {Abraham}}, \bibinfo {author} {\bibfnamefont {S.}~\bibnamefont {Danieli}}, \bibinfo {author} {\bibfnamefont {A.~J.}\ \bibnamefont {Romanowsky}}, \ and\ \bibinfo {author} {\bibfnamefont {G.}~\bibnamefont {Tremblay}},\ }\href {\doibase 10.3847/2041-8213/ae3d0e} {\bibfield  {journal} {\bibinfo  {journal} {Astrophys. J. Lett.}\ }\textbf {\bibinfo {volume} {998}},\ \bibinfo {pages} {L27} (\bibinfo {year} {2026})},\ \Eprint {http://arxiv.org/abs/2512.04166} {arXiv:2512.04166 [astro-ph.GA]} \BibitemShut {NoStop}%
\bibitem [{\citenamefont {{Lau}}\ and\ \citenamefont {{Lai}}(2026)}]{slowlenses:floaters:2026arXiv260212017L}%
  \BibitemOpen
  \bibfield  {author} {\bibinfo {author} {\bibfnamefont {J.~Y.}\ \bibnamefont {{Lau}}}\ and\ \bibinfo {author} {\bibfnamefont {D.}~\bibnamefont {{Lai}}},\ }\href {\doibase 10.48550/arXiv.2602.12017} {\bibfield  {journal} {\bibinfo  {journal} {arXiv e-prints}\ ,\ \bibinfo {eid} {arXiv:2602.12017}} (\bibinfo {year} {2026})},\ \Eprint {http://arxiv.org/abs/2602.12017} {arXiv:2602.12017 [astro-ph.EP]} \BibitemShut {NoStop}%
\bibitem [{\citenamefont {{Dong, et al.}}(2026)}]{slowlenses:FFPOGLE:2026Science}%
  \BibitemOpen
  \bibfield  {author} {\bibinfo {author} {\bibfnamefont {S.}~\bibnamefont {{Dong, et al.}}},\ }\href {\doibase 10.1126/science.adv9266} {\bibfield  {journal} {\bibinfo  {journal} {Science}\ }\textbf {\bibinfo {volume} {391}},\ \bibinfo {pages} {96} (\bibinfo {year} {2026})},\ \Eprint {http://arxiv.org/abs/2601.00057} {arXiv:2601.00057 [astro-ph.EP]} \BibitemShut {NoStop}%
\bibitem [{\citenamefont {{Bottema}}(1993)}]{slowlenses:diskformed1993}%
  \BibitemOpen
  \bibfield  {author} {\bibinfo {author} {\bibfnamefont {R.}~\bibnamefont {{Bottema}}},\ }\href@noop {} {\bibfield  {journal} {\bibinfo  {journal} {AAP}\ }\textbf {\bibinfo {volume} {275}},\ \bibinfo {pages} {16} (\bibinfo {year} {1993})}\BibitemShut {NoStop}%
\bibitem [{\citenamefont {{Purcell}}\ \emph {et~al.}(2009)\citenamefont {{Purcell}}, \citenamefont {{Bullock}},\ and\ \citenamefont {{Kaplinghat}}}]{slowlenses:darkdiskkaplinghat:2009ApJ}%
  \BibitemOpen
  \bibfield  {author} {\bibinfo {author} {\bibfnamefont {C.~W.}\ \bibnamefont {{Purcell}}}, \bibinfo {author} {\bibfnamefont {J.~S.}\ \bibnamefont {{Bullock}}}, \ and\ \bibinfo {author} {\bibfnamefont {M.}~\bibnamefont {{Kaplinghat}}},\ }\href {\doibase 10.1088/0004-637X/703/2/2275} {\bibfield  {journal} {\bibinfo  {journal} {APJ}\ }\textbf {\bibinfo {volume} {703}},\ \bibinfo {pages} {2275} (\bibinfo {year} {2009})},\ \Eprint {http://arxiv.org/abs/0906.5348} {arXiv:0906.5348 [astro-ph.GA]} \BibitemShut {NoStop}%
\bibitem [{\citenamefont {Winch}\ \emph {et~al.}(2022)\citenamefont {Winch}, \citenamefont {Setford}, \citenamefont {Bovy},\ and\ \citenamefont {Curtin}}]{darkdisk:WinchSetfordCurtin:2020cju}%
  \BibitemOpen
  \bibfield  {author} {\bibinfo {author} {\bibfnamefont {H.}~\bibnamefont {Winch}}, \bibinfo {author} {\bibfnamefont {J.}~\bibnamefont {Setford}}, \bibinfo {author} {\bibfnamefont {J.}~\bibnamefont {Bovy}}, \ and\ \bibinfo {author} {\bibfnamefont {D.}~\bibnamefont {Curtin}},\ }\href {\doibase 10.3847/1538-4357/ac7467} {\bibfield  {journal} {\bibinfo  {journal} {Astrophys. J.}\ }\textbf {\bibinfo {volume} {933}},\ \bibinfo {pages} {177} (\bibinfo {year} {2022})},\ \Eprint {http://arxiv.org/abs/2012.07136} {arXiv:2012.07136 [astro-ph.GA]} \BibitemShut {NoStop}%
\bibitem [{\citenamefont {Buckley}\ and\ \citenamefont {DiFranzo}(2018)}]{fragmented:Buckley:2017ttd}%
  \BibitemOpen
  \bibfield  {author} {\bibinfo {author} {\bibfnamefont {M.~R.}\ \bibnamefont {Buckley}}\ and\ \bibinfo {author} {\bibfnamefont {A.}~\bibnamefont {DiFranzo}},\ }\href {\doibase 10.1103/PhysRevLett.120.051102} {\bibfield  {journal} {\bibinfo  {journal} {Phys. Rev. Lett.}\ }\textbf {\bibinfo {volume} {120}},\ \bibinfo {pages} {051102} (\bibinfo {year} {2018})},\ \Eprint {http://arxiv.org/abs/1707.03829} {arXiv:1707.03829 [hep-ph]} \BibitemShut {NoStop}%
\bibitem [{\citenamefont {D'Amico}\ \emph {et~al.}(2018)\citenamefont {D'Amico}, \citenamefont {Panci}, \citenamefont {Lupi}, \citenamefont {Bovino},\ and\ \citenamefont {Silk}}]{fragmented:DAmico:2017lqj}%
  \BibitemOpen
  \bibfield  {author} {\bibinfo {author} {\bibfnamefont {G.}~\bibnamefont {D'Amico}}, \bibinfo {author} {\bibfnamefont {P.}~\bibnamefont {Panci}}, \bibinfo {author} {\bibfnamefont {A.}~\bibnamefont {Lupi}}, \bibinfo {author} {\bibfnamefont {S.}~\bibnamefont {Bovino}}, \ and\ \bibinfo {author} {\bibfnamefont {J.}~\bibnamefont {Silk}},\ }\href {\doibase 10.1093/mnras/stx2419} {\bibfield  {journal} {\bibinfo  {journal} {Mon. Not. Roy. Astron. Soc.}\ }\textbf {\bibinfo {volume} {473}},\ \bibinfo {pages} {328} (\bibinfo {year} {2018})},\ \Eprint {http://arxiv.org/abs/1707.03419} {arXiv:1707.03419 [astro-ph.CO]} \BibitemShut {NoStop}%
\bibitem [{\citenamefont {Chang}\ \emph {et~al.}(2019)\citenamefont {Chang}, \citenamefont {Egana-Ugrinovic}, \citenamefont {Essig},\ and\ \citenamefont {Kouvaris}}]{fragmented:ChangDanielEssigKouvaris:2018bgx}%
  \BibitemOpen
  \bibfield  {author} {\bibinfo {author} {\bibfnamefont {J.~H.}\ \bibnamefont {Chang}}, \bibinfo {author} {\bibfnamefont {D.}~\bibnamefont {Egana-Ugrinovic}}, \bibinfo {author} {\bibfnamefont {R.}~\bibnamefont {Essig}}, \ and\ \bibinfo {author} {\bibfnamefont {C.}~\bibnamefont {Kouvaris}},\ }\href {\doibase 10.1088/1475-7516/2019/03/036} {\bibfield  {journal} {\bibinfo  {journal} {JCAP}\ }\textbf {\bibinfo {volume} {03}},\ \bibinfo {pages} {036} (\bibinfo {year} {2019})},\ \Eprint {http://arxiv.org/abs/1812.07000} {arXiv:1812.07000 [hep-ph]} \BibitemShut {NoStop}%
\bibitem [{\citenamefont {Gurian}\ \emph {et~al.}(2022)\citenamefont {Gurian}, \citenamefont {Ryan}, \citenamefont {Schon}, \citenamefont {Jeong},\ and\ \citenamefont {Shandera}}]{fragmented:Gurian:2022nbx}%
  \BibitemOpen
  \bibfield  {author} {\bibinfo {author} {\bibfnamefont {J.}~\bibnamefont {Gurian}}, \bibinfo {author} {\bibfnamefont {M.}~\bibnamefont {Ryan}}, \bibinfo {author} {\bibfnamefont {S.}~\bibnamefont {Schon}}, \bibinfo {author} {\bibfnamefont {D.}~\bibnamefont {Jeong}}, \ and\ \bibinfo {author} {\bibfnamefont {S.}~\bibnamefont {Shandera}},\ }\href {\doibase 10.3847/2041-8213/ac997c} {\bibfield  {journal} {\bibinfo  {journal} {Astrophys. J. Lett.}\ }\textbf {\bibinfo {volume} {939}},\ \bibinfo {pages} {L12} (\bibinfo {year} {2022})},\ \bibinfo {note} {[Erratum: Astrophys.J.Lett. 949, L44 (2023), Erratum: Astrophys.J. 949, L44 (2023)]},\ \Eprint {http://arxiv.org/abs/2209.00064} {arXiv:2209.00064 [astro-ph.CO]} \BibitemShut {NoStop}%
\bibitem [{\citenamefont {Roy}\ \emph {et~al.}(2023)\citenamefont {Roy}, \citenamefont {Shen}, \citenamefont {Lisanti}, \citenamefont {Curtin}, \citenamefont {Murray},\ and\ \citenamefont {Hopkins}}]{fragmented:LisantiCurtin:2023zar}%
  \BibitemOpen
  \bibfield  {author} {\bibinfo {author} {\bibfnamefont {S.}~\bibnamefont {Roy}}, \bibinfo {author} {\bibfnamefont {X.}~\bibnamefont {Shen}}, \bibinfo {author} {\bibfnamefont {M.}~\bibnamefont {Lisanti}}, \bibinfo {author} {\bibfnamefont {D.}~\bibnamefont {Curtin}}, \bibinfo {author} {\bibfnamefont {N.}~\bibnamefont {Murray}}, \ and\ \bibinfo {author} {\bibfnamefont {P.~F.}\ \bibnamefont {Hopkins}},\ }\href {\doibase 10.3847/2041-8213/ace2c8} {\bibfield  {journal} {\bibinfo  {journal} {Astrophys. J. Lett.}\ }\textbf {\bibinfo {volume} {954}},\ \bibinfo {pages} {L40} (\bibinfo {year} {2023})},\ \Eprint {http://arxiv.org/abs/2304.09878} {arXiv:2304.09878 [astro-ph.GA]} \BibitemShut {NoStop}%
\bibitem [{\citenamefont {Bramante}\ \emph {et~al.}(2024)\citenamefont {Bramante}, \citenamefont {Diamond},\ and\ \citenamefont {Kim}}]{fragmented:Bramante:2023ddr}%
  \BibitemOpen
  \bibfield  {author} {\bibinfo {author} {\bibfnamefont {J.}~\bibnamefont {Bramante}}, \bibinfo {author} {\bibfnamefont {M.}~\bibnamefont {Diamond}}, \ and\ \bibinfo {author} {\bibfnamefont {J.~L.}\ \bibnamefont {Kim}},\ }\href {\doibase 10.1088/1475-7516/2024/02/002} {\bibfield  {journal} {\bibinfo  {journal} {JCAP}\ }\textbf {\bibinfo {volume} {02}},\ \bibinfo {pages} {002} (\bibinfo {year} {2024})},\ \Eprint {http://arxiv.org/abs/2309.13148} {arXiv:2309.13148 [hep-ph]} \BibitemShut {NoStop}%
\bibitem [{\citenamefont {Osuna}\ and\ \citenamefont {Shandera}(2026)}]{fragmented:Osuna:2026dzj}%
  \BibitemOpen
  \bibfield  {author} {\bibinfo {author} {\bibfnamefont {J.~C.}\ \bibnamefont {Osuna}}\ and\ \bibinfo {author} {\bibfnamefont {S.}~\bibnamefont {Shandera}},\ }\href@noop {} {\  (\bibinfo {year} {2026})},\ \Eprint {http://arxiv.org/abs/2603.24862} {arXiv:2603.24862 [astro-ph.CO]} \BibitemShut {NoStop}%
\bibitem [{\citenamefont {{Gould}}(1992)}]{parallaxgould1992ApJ}%
  \BibitemOpen
  \bibfield  {author} {\bibinfo {author} {\bibfnamefont {A.}~\bibnamefont {{Gould}}},\ }\href {\doibase 10.1086/171443} {\bibfield  {journal} {\bibinfo  {journal} {APJ}\ }\textbf {\bibinfo {volume} {392}},\ \bibinfo {pages} {442} (\bibinfo {year} {1992})}\BibitemShut {NoStop}%
\bibitem [{\citenamefont {{Alcock}}\ \emph {et~al.}(1995)\citenamefont {{Alcock}}, \citenamefont {{Allsman}}, \citenamefont {{Alves}}, \citenamefont {{Axelrod}}, \citenamefont {{Bennett}}, \citenamefont {{Cook}}, \citenamefont {{Freeman}}, \citenamefont {{Griest}}, \citenamefont {{Guern}}, \citenamefont {{Lehner}}, \citenamefont {{Marshall}}, \citenamefont {{Peterson}}, \citenamefont {{Pratt}}, \citenamefont {{Quinn}}, \citenamefont {{Rodgers}}, \citenamefont {{Stubbs}},\ and\ \citenamefont {{Sutherland}}}]{parallaxalcock1995ApJ}%
  \BibitemOpen
  \bibfield  {author} {\bibinfo {author} {\bibfnamefont {C.}~\bibnamefont {{Alcock}}}, \bibinfo {author} {\bibfnamefont {R.~A.}\ \bibnamefont {{Allsman}}}, \bibinfo {author} {\bibfnamefont {D.}~\bibnamefont {{Alves}}}, \bibinfo {author} {\bibfnamefont {T.~S.}\ \bibnamefont {{Axelrod}}}, \bibinfo {author} {\bibfnamefont {D.~P.}\ \bibnamefont {{Bennett}}}, \bibinfo {author} {\bibfnamefont {K.~H.}\ \bibnamefont {{Cook}}}, \bibinfo {author} {\bibfnamefont {K.~C.}\ \bibnamefont {{Freeman}}}, \bibinfo {author} {\bibfnamefont {K.}~\bibnamefont {{Griest}}}, \bibinfo {author} {\bibfnamefont {J.}~\bibnamefont {{Guern}}}, \bibinfo {author} {\bibfnamefont {M.~J.}\ \bibnamefont {{Lehner}}}, \bibinfo {author} {\bibfnamefont {S.~L.}\ \bibnamefont {{Marshall}}}, \bibinfo {author} {\bibfnamefont {B.~A.}\ \bibnamefont {{Peterson}}}, \bibinfo {author} {\bibfnamefont {M.~R.}\ \bibnamefont {{Pratt}}}, \bibinfo {author} {\bibfnamefont {P.~J.}\ \bibnamefont {{Quinn}}}, \bibinfo {author} {\bibfnamefont {A.~W.}\ \bibnamefont
  {{Rodgers}}}, \bibinfo {author} {\bibfnamefont {C.~W.}\ \bibnamefont {{Stubbs}}}, \ and\ \bibinfo {author} {\bibfnamefont {W.}~\bibnamefont {{Sutherland}}},\ }\href {\doibase 10.1086/309783} {\bibfield  {journal} {\bibinfo  {journal} {APJL}\ }\textbf {\bibinfo {volume} {454}},\ \bibinfo {pages} {L125} (\bibinfo {year} {1995})},\ \Eprint {http://arxiv.org/abs/astro-ph/9506114} {arXiv:astro-ph/9506114 [astro-ph]} \BibitemShut {NoStop}%
\bibitem [{\citenamefont {{Hog}}\ \emph {et~al.}(1995)\citenamefont {{Hog}}, \citenamefont {{Novikov}},\ and\ \citenamefont {{Polnarev}}}]{1995A&A...294..287H}%
  \BibitemOpen
  \bibfield  {author} {\bibinfo {author} {\bibfnamefont {E.}~\bibnamefont {{Hog}}}, \bibinfo {author} {\bibfnamefont {I.~D.}\ \bibnamefont {{Novikov}}}, \ and\ \bibinfo {author} {\bibfnamefont {A.~G.}\ \bibnamefont {{Polnarev}}},\ }\href@noop {} {\bibfield  {journal} {\bibinfo  {journal} {\aap}\ }\textbf {\bibinfo {volume} {294}},\ \bibinfo {pages} {287} (\bibinfo {year} {1995})}\BibitemShut {NoStop}%
\bibitem [{\citenamefont {{Lu}}\ \emph {et~al.}(2016)\citenamefont {{Lu}}, \citenamefont {{Sinukoff}}, \citenamefont {{Ofek}}, \citenamefont {{Udalski}},\ and\ \citenamefont {{Kozlowski}}}]{2016ApJ...830...41L}%
  \BibitemOpen
  \bibfield  {author} {\bibinfo {author} {\bibfnamefont {J.~R.}\ \bibnamefont {{Lu}}}, \bibinfo {author} {\bibfnamefont {E.}~\bibnamefont {{Sinukoff}}}, \bibinfo {author} {\bibfnamefont {E.~O.}\ \bibnamefont {{Ofek}}}, \bibinfo {author} {\bibfnamefont {A.}~\bibnamefont {{Udalski}}}, \ and\ \bibinfo {author} {\bibfnamefont {S.}~\bibnamefont {{Kozlowski}}},\ }\href {\doibase 10.3847/0004-637X/830/1/41} {\bibfield  {journal} {\bibinfo  {journal} {\apj}\ }\textbf {\bibinfo {volume} {830}},\ \bibinfo {eid} {41} (\bibinfo {year} {2016})},\ \Eprint {http://arxiv.org/abs/1607.08284} {arXiv:1607.08284 [astro-ph.SR]} \BibitemShut {NoStop}%
\bibitem [{\citenamefont {{Lee}}(2017)}]{2017Univ....3...53L}%
  \BibitemOpen
  \bibfield  {author} {\bibinfo {author} {\bibfnamefont {C.-H.}\ \bibnamefont {{Lee}}},\ }\href {\doibase 10.3390/universe3030053} {\bibfield  {journal} {\bibinfo  {journal} {Universe}\ }\textbf {\bibinfo {volume} {3}},\ \bibinfo {eid} {53} (\bibinfo {year} {2017})},\ \Eprint {http://arxiv.org/abs/1711.05298} {arXiv:1711.05298 [astro-ph.IM]} \BibitemShut {NoStop}%
\bibitem [{\citenamefont {{Sahu, et al.}}\ and\ \citenamefont {{RoboNet Collaboration}}(2022)}]{2022ApJ...933...83S}%
  \BibitemOpen
  \bibfield  {author} {\bibinfo {author} {\bibfnamefont {K.~C.}\ \bibnamefont {{Sahu, et al.}}}\ and\ \bibinfo {author} {\bibnamefont {{RoboNet Collaboration}}},\ }\href {\doibase 10.3847/1538-4357/ac739e} {\bibfield  {journal} {\bibinfo  {journal} {\apj}\ }\textbf {\bibinfo {volume} {933}},\ \bibinfo {eid} {83} (\bibinfo {year} {2022})},\ \Eprint {http://arxiv.org/abs/2201.13296} {arXiv:2201.13296 [astro-ph.SR]} \BibitemShut {NoStop}%
\bibitem [{\citenamefont {{Lam, et al.}}(2022)}]{2022ApJ...933L..23L}%
  \BibitemOpen
  \bibfield  {author} {\bibinfo {author} {\bibfnamefont {C.~Y.}\ \bibnamefont {{Lam, et al.}}},\ }\href {\doibase 10.3847/2041-8213/ac7442} {\bibfield  {journal} {\bibinfo  {journal} {\apjl}\ }\textbf {\bibinfo {volume} {933}},\ \bibinfo {eid} {L23} (\bibinfo {year} {2022})},\ \Eprint {http://arxiv.org/abs/2202.01903} {arXiv:2202.01903 [astro-ph.GA]} \BibitemShut {NoStop}%
\bibitem [{\citenamefont {{Griest}}(1991)}]{griest1991ApJ}%
  \BibitemOpen
  \bibfield  {author} {\bibinfo {author} {\bibfnamefont {K.}~\bibnamefont {{Griest}}},\ }\href {\doibase 10.1086/169575} {\bibfield  {journal} {\bibinfo  {journal} {APJ}\ }\textbf {\bibinfo {volume} {366}},\ \bibinfo {pages} {412} (\bibinfo {year} {1991})}\BibitemShut {NoStop}%
\bibitem [{\citenamefont {van~der Marel}\ \emph {et~al.}(2012)\citenamefont {van~der Marel}, \citenamefont {Fardal}, \citenamefont {Besla}, \citenamefont {Beaton}, \citenamefont {Sohn}, \citenamefont {Anderson}, \citenamefont {Brown},\ and\ \citenamefont {Guhathakurta}}]{M31velocity:vanderMarel2012}%
  \BibitemOpen
  \bibfield  {author} {\bibinfo {author} {\bibfnamefont {R.~P.}\ \bibnamefont {van~der Marel}}, \bibinfo {author} {\bibfnamefont {M.}~\bibnamefont {Fardal}}, \bibinfo {author} {\bibfnamefont {G.}~\bibnamefont {Besla}}, \bibinfo {author} {\bibfnamefont {R.~L.}\ \bibnamefont {Beaton}}, \bibinfo {author} {\bibfnamefont {S.~T.}\ \bibnamefont {Sohn}}, \bibinfo {author} {\bibfnamefont {J.}~\bibnamefont {Anderson}}, \bibinfo {author} {\bibfnamefont {T.}~\bibnamefont {Brown}}, \ and\ \bibinfo {author} {\bibfnamefont {P.}~\bibnamefont {Guhathakurta}},\ }\href {\doibase 10.1088/0004-637x/753/1/8} {\bibfield  {journal} {\bibinfo  {journal} {The Astrophysical Journal}\ }\textbf {\bibinfo {volume} {753}},\ \bibinfo {pages} {8} (\bibinfo {year} {2012})}\BibitemShut {NoStop}%
\bibitem [{\citenamefont {Niikura}\ \emph {et~al.}(2019{\natexlab{b}})\citenamefont {Niikura}, \citenamefont {Takada}, \citenamefont {Yokoyama}, \citenamefont {Sumi},\ and\ \citenamefont {Masaki}}]{Niikura:2019kqi}%
  \BibitemOpen
  \bibfield  {author} {\bibinfo {author} {\bibfnamefont {H.}~\bibnamefont {Niikura}}, \bibinfo {author} {\bibfnamefont {M.}~\bibnamefont {Takada}}, \bibinfo {author} {\bibfnamefont {S.}~\bibnamefont {Yokoyama}}, \bibinfo {author} {\bibfnamefont {T.}~\bibnamefont {Sumi}}, \ and\ \bibinfo {author} {\bibfnamefont {S.}~\bibnamefont {Masaki}},\ }\href {\doibase 10.1103/PhysRevD.99.083503} {\bibfield  {journal} {\bibinfo  {journal} {Phys. Rev. D}\ }\textbf {\bibinfo {volume} {99}},\ \bibinfo {pages} {083503} (\bibinfo {year} {2019}{\natexlab{b}})},\ \Eprint {http://arxiv.org/abs/1901.07120} {arXiv:1901.07120 [astro-ph.CO]} \BibitemShut {NoStop}%
\bibitem [{\citenamefont {Green}(2026)}]{review:Green:2026xhw}%
  \BibitemOpen
  \bibfield  {author} {\bibinfo {author} {\bibfnamefont {A.~M.}\ \bibnamefont {Green}},\ }\href@noop {} {\  (\bibinfo {year} {2026})},\ \Eprint {http://arxiv.org/abs/2602.15974} {arXiv:2602.15974 [astro-ph.GA]} \BibitemShut {NoStop}%
\bibitem [{\citenamefont {{Witt}}\ and\ \citenamefont {{Mao}}(1994)}]{wittmao1994}%
  \BibitemOpen
  \bibfield  {author} {\bibinfo {author} {\bibfnamefont {H.~J.}\ \bibnamefont {{Witt}}}\ and\ \bibinfo {author} {\bibfnamefont {S.}~\bibnamefont {{Mao}}},\ }\href {\doibase 10.1086/174426} {\bibfield  {journal} {\bibinfo  {journal} {APJ}\ }\textbf {\bibinfo {volume} {430}},\ \bibinfo {pages} {505} (\bibinfo {year} {1994})}\BibitemShut {NoStop}%
\bibitem [{\citenamefont {Smyth}\ \emph {et~al.}(2020)\citenamefont {Smyth}, \citenamefont {Profumo}, \citenamefont {English}, \citenamefont {Jeltema}, \citenamefont {McKinnon},\ and\ \citenamefont {Guhathakurta}}]{Smyth:2019whb}%
  \BibitemOpen
  \bibfield  {author} {\bibinfo {author} {\bibfnamefont {N.}~\bibnamefont {Smyth}}, \bibinfo {author} {\bibfnamefont {S.}~\bibnamefont {Profumo}}, \bibinfo {author} {\bibfnamefont {S.}~\bibnamefont {English}}, \bibinfo {author} {\bibfnamefont {T.}~\bibnamefont {Jeltema}}, \bibinfo {author} {\bibfnamefont {K.}~\bibnamefont {McKinnon}}, \ and\ \bibinfo {author} {\bibfnamefont {P.}~\bibnamefont {Guhathakurta}},\ }\href {\doibase 10.1103/PhysRevD.101.063005} {\bibfield  {journal} {\bibinfo  {journal} {Phys. Rev. D}\ }\textbf {\bibinfo {volume} {101}},\ \bibinfo {pages} {063005} (\bibinfo {year} {2020})},\ \Eprint {http://arxiv.org/abs/1910.01285} {arXiv:1910.01285 [astro-ph.CO]} \BibitemShut {NoStop}%
\bibitem [{\citenamefont {Baltz}\ and\ \citenamefont {Silk}(2000)}]{BaltzSilk:1999fr}%
  \BibitemOpen
  \bibfield  {author} {\bibinfo {author} {\bibfnamefont {E.~A.}\ \bibnamefont {Baltz}}\ and\ \bibinfo {author} {\bibfnamefont {J.}~\bibnamefont {Silk}},\ }\href {\doibase 10.1086/308385} {\bibfield  {journal} {\bibinfo  {journal} {Astrophys. J.}\ }\textbf {\bibinfo {volume} {530}},\ \bibinfo {pages} {578} (\bibinfo {year} {2000})},\ \Eprint {http://arxiv.org/abs/astro-ph/9901408} {arXiv:astro-ph/9901408} \BibitemShut {NoStop}%
\bibitem [{\citenamefont {Sugiyama}\ \emph {et~al.}(2026)\citenamefont {Sugiyama}, \citenamefont {Takada}, \citenamefont {Yasuda},\ and\ \citenamefont {Tominaga}}]{SubaruHSC:Sugiyama:2026kpv}%
  \BibitemOpen
  \bibfield  {author} {\bibinfo {author} {\bibfnamefont {S.}~\bibnamefont {Sugiyama}}, \bibinfo {author} {\bibfnamefont {M.}~\bibnamefont {Takada}}, \bibinfo {author} {\bibfnamefont {N.}~\bibnamefont {Yasuda}}, \ and\ \bibinfo {author} {\bibfnamefont {N.}~\bibnamefont {Tominaga}},\ }\href@noop {} {\  (\bibinfo {year} {2026})},\ \Eprint {http://arxiv.org/abs/2602.05840} {arXiv:2602.05840 [astro-ph.CO]} \BibitemShut {NoStop}%
\bibitem [{\citenamefont {Mr{\'o}z}\ and\ \citenamefont {Udalski}(2026)}]{SubaruHSC:Mroz:2026nez}%
  \BibitemOpen
  \bibfield  {author} {\bibinfo {author} {\bibfnamefont {P.}~\bibnamefont {Mr{\'o}z}}\ and\ \bibinfo {author} {\bibfnamefont {A.}~\bibnamefont {Udalski}},\ }\href@noop {} {\  (\bibinfo {year} {2026})},\ \Eprint {http://arxiv.org/abs/2604.00111} {arXiv:2604.00111 [astro-ph.CO]} \BibitemShut {NoStop}%
\bibitem [{Ver()}]{VermaTamtaRajhaloindep}%
  \BibitemOpen
  \href@noop {} {}\bibinfo {howpublished} {H. Verma, M. Tamta, N. Raj. In preparation.}\BibitemShut {Stop}%
\bibitem [{\citenamefont {Katz}\ \emph {et~al.}(2018)\citenamefont {Katz}, \citenamefont {Kopp}, \citenamefont {Sibiryakov},\ and\ \citenamefont {Xue}}]{femtolensing:Katz:2018JCAP}%
  \BibitemOpen
  \bibfield  {author} {\bibinfo {author} {\bibfnamefont {A.}~\bibnamefont {Katz}}, \bibinfo {author} {\bibfnamefont {J.}~\bibnamefont {Kopp}}, \bibinfo {author} {\bibfnamefont {S.}~\bibnamefont {Sibiryakov}}, \ and\ \bibinfo {author} {\bibfnamefont {W.}~\bibnamefont {Xue}},\ }\href {\doibase 10.1088/1475-7516/2018/12/005} {\bibfield  {journal} {\bibinfo  {journal} {Journal of Cosmology and Astroparticle Physics}\ }\textbf {\bibinfo {volume} {2018}},\ \bibinfo {pages} {005} (\bibinfo {year} {2018})},\ \Eprint {http://arxiv.org/abs/1807.11495} {arXiv:1807.11495} \BibitemShut {NoStop}%
\bibitem [{\citenamefont {Jung}\ and\ \citenamefont {Kim}(2020)}]{femtolensing:Jung:2020PRR}%
  \BibitemOpen
  \bibfield  {author} {\bibinfo {author} {\bibfnamefont {S.}~\bibnamefont {Jung}}\ and\ \bibinfo {author} {\bibfnamefont {T.}~\bibnamefont {Kim}},\ }\href {\doibase 10.1103/PhysRevResearch.2.013113} {\bibfield  {journal} {\bibinfo  {journal} {Physical Review Research}\ }\textbf {\bibinfo {volume} {2}},\ \bibinfo {pages} {013113} (\bibinfo {year} {2020})},\ \Eprint {http://arxiv.org/abs/1908.00078} {arXiv:1908.00078} \BibitemShut {NoStop}%
\bibitem [{\citenamefont {Gawade}\ \emph {et~al.}(2023)\citenamefont {Gawade}, \citenamefont {More},\ and\ \citenamefont {Bhalerao}}]{picolensingDaksha:Gawade:2023gmt}%
  \BibitemOpen
  \bibfield  {author} {\bibinfo {author} {\bibfnamefont {P.}~\bibnamefont {Gawade}}, \bibinfo {author} {\bibfnamefont {S.}~\bibnamefont {More}}, \ and\ \bibinfo {author} {\bibfnamefont {V.}~\bibnamefont {Bhalerao}},\ }\href {\doibase 10.1093/mnras/stad3336} {\bibfield  {journal} {\bibinfo  {journal} {Mon. Not. Roy. Astron. Soc.}\ }\textbf {\bibinfo {volume} {527}},\ \bibinfo {pages} {3306} (\bibinfo {year} {2023})},\ \Eprint {http://arxiv.org/abs/2308.01775} {arXiv:2308.01775 [astro-ph.CO]} \BibitemShut {NoStop}%
\bibitem [{\citenamefont {Fedderke}\ and\ \citenamefont {Sibiryakov}(2025)}]{picolensing:Fedderke:2025PRD}%
  \BibitemOpen
  \bibfield  {author} {\bibinfo {author} {\bibfnamefont {M.~A.}\ \bibnamefont {Fedderke}}\ and\ \bibinfo {author} {\bibfnamefont {S.}~\bibnamefont {Sibiryakov}},\ }\href {\doibase 10.1103/PhysRevD.111.063060} {\bibfield  {journal} {\bibinfo  {journal} {Physical Review D}\ }\textbf {\bibinfo {volume} {111}},\ \bibinfo {pages} {063060} (\bibinfo {year} {2025})},\ \Eprint {http://arxiv.org/abs/2411.12947} {arXiv:2411.12947} \BibitemShut {NoStop}%
\bibitem [{\citenamefont {Drlica-Wagner}\ \emph {et~al.}(2019)\citenamefont {Drlica-Wagner} \emph {et~al.}}]{RubinreachDrlicaWagner:2019mwo}%
  \BibitemOpen
  \bibfield  {author} {\bibinfo {author} {\bibfnamefont {A.}~\bibnamefont {Drlica-Wagner}} \emph {et~al.} (\bibinfo {collaboration} {LSST Dark Matter Group}),\ }\href@noop {} {\  (\bibinfo {year} {2019})},\ \Eprint {http://arxiv.org/abs/1902.01055} {arXiv:1902.01055 [astro-ph.CO]} \BibitemShut {NoStop}%
\bibitem [{\citenamefont {Crispim~Romao}\ \emph {et~al.}(2025)\citenamefont {Crispim~Romao}, \citenamefont {Croon}, \citenamefont {Crossey},\ and\ \citenamefont {Godines}}]{RubinreachCroonRomao:2025kxx}%
  \BibitemOpen
  \bibfield  {author} {\bibinfo {author} {\bibfnamefont {M.}~\bibnamefont {Crispim~Romao}}, \bibinfo {author} {\bibfnamefont {D.}~\bibnamefont {Croon}}, \bibinfo {author} {\bibfnamefont {B.}~\bibnamefont {Crossey}}, \ and\ \bibinfo {author} {\bibfnamefont {D.}~\bibnamefont {Godines}},\ }\href {\doibase 10.1088/1475-7516/2025/10/066} {\bibfield  {journal} {\bibinfo  {journal} {JCAP}\ }\textbf {\bibinfo {volume} {10}},\ \bibinfo {pages} {066} (\bibinfo {year} {2025})},\ \Eprint {http://arxiv.org/abs/2506.20709} {arXiv:2506.20709 [astro-ph.CO]} \BibitemShut {NoStop}%
\bibitem [{\citenamefont {DeRocco}\ \emph {et~al.}(2024)\citenamefont {DeRocco}, \citenamefont {Frangipane}, \citenamefont {Hamer}, \citenamefont {Profumo},\ and\ \citenamefont {Smyth}}]{Romanreach:DeRoccoProfumoSmyth:2023gde}%
  \BibitemOpen
  \bibfield  {author} {\bibinfo {author} {\bibfnamefont {W.}~\bibnamefont {DeRocco}}, \bibinfo {author} {\bibfnamefont {E.}~\bibnamefont {Frangipane}}, \bibinfo {author} {\bibfnamefont {N.}~\bibnamefont {Hamer}}, \bibinfo {author} {\bibfnamefont {S.}~\bibnamefont {Profumo}}, \ and\ \bibinfo {author} {\bibfnamefont {N.}~\bibnamefont {Smyth}},\ }\href {\doibase 10.1103/PhysRevD.109.023013} {\bibfield  {journal} {\bibinfo  {journal} {Phys. Rev. D}\ }\textbf {\bibinfo {volume} {109}},\ \bibinfo {pages} {023013} (\bibinfo {year} {2024})},\ \Eprint {http://arxiv.org/abs/2311.00751} {arXiv:2311.00751 [astro-ph.CO]} \BibitemShut {NoStop}%
\bibitem [{\citenamefont {Fardeen}\ \emph {et~al.}(2024)\citenamefont {Fardeen}, \citenamefont {McGill}, \citenamefont {Perkins}, \citenamefont {Dawson}, \citenamefont {Abrams}, \citenamefont {Lu}, \citenamefont {Ho},\ and\ \citenamefont {Bird}}]{Romanreachastrometric:fardeen2024}%
  \BibitemOpen
  \bibfield  {author} {\bibinfo {author} {\bibfnamefont {J.}~\bibnamefont {Fardeen}}, \bibinfo {author} {\bibfnamefont {P.}~\bibnamefont {McGill}}, \bibinfo {author} {\bibfnamefont {S.~E.}\ \bibnamefont {Perkins}}, \bibinfo {author} {\bibfnamefont {W.~A.}\ \bibnamefont {Dawson}}, \bibinfo {author} {\bibfnamefont {N.~S.}\ \bibnamefont {Abrams}}, \bibinfo {author} {\bibfnamefont {J.~R.}\ \bibnamefont {Lu}}, \bibinfo {author} {\bibfnamefont {M.-F.}\ \bibnamefont {Ho}}, \ and\ \bibinfo {author} {\bibfnamefont {S.}~\bibnamefont {Bird}},\ }\href {https://arxiv.org/abs/2312.13249} {\enquote {\bibinfo {title} {Astrometric microlensing by primordial black holes with the roman space telescope},}\ } (\bibinfo {year} {2024}),\ \Eprint {http://arxiv.org/abs/2312.13249} {arXiv:2312.13249 [astro-ph.GA]} \BibitemShut {NoStop}%
\bibitem [{\citenamefont {Hamolli}\ \emph {et~al.}(2021)\citenamefont {Hamolli}, \citenamefont {Hafizi}, \citenamefont {De~Paolis},\ and\ \citenamefont {Nucita}}]{Euclidreach:Hamolli2021}%
  \BibitemOpen
  \bibfield  {author} {\bibinfo {author} {\bibfnamefont {L.}~\bibnamefont {Hamolli}}, \bibinfo {author} {\bibfnamefont {M.}~\bibnamefont {Hafizi}}, \bibinfo {author} {\bibfnamefont {F.}~\bibnamefont {De~Paolis}}, \ and\ \bibinfo {author} {\bibfnamefont {A.~A.}\ \bibnamefont {Nucita}},\ }\href {\doibase 10.1007/s10509-021-03980-0} {\bibfield  {journal} {\bibinfo  {journal} {Astrophysics and Space Science}\ }\textbf {\bibinfo {volume} {366}} (\bibinfo {year} {2021}),\ 10.1007/s10509-021-03980-0}\BibitemShut {NoStop}%
\bibitem [{\citenamefont {Bai}\ and\ \citenamefont {Orlofsky}(2019)}]{xraymicrolensing:Bai:2018bej}%
  \BibitemOpen
  \bibfield  {author} {\bibinfo {author} {\bibfnamefont {Y.}~\bibnamefont {Bai}}\ and\ \bibinfo {author} {\bibfnamefont {N.}~\bibnamefont {Orlofsky}},\ }\href {\doibase 10.1103/PhysRevD.99.123019} {\bibfield  {journal} {\bibinfo  {journal} {Phys. Rev. D}\ }\textbf {\bibinfo {volume} {99}},\ \bibinfo {pages} {123019} (\bibinfo {year} {2019})},\ \Eprint {http://arxiv.org/abs/1812.01427} {arXiv:1812.01427 [astro-ph.HE]} \BibitemShut {NoStop}%
\bibitem [{\citenamefont {Tamta}\ \emph {et~al.}(2025)\citenamefont {Tamta}, \citenamefont {Raj},\ and\ \citenamefont {Sharma}}]{xraymicrolensing:Tamta:2024pow}%
  \BibitemOpen
  \bibfield  {author} {\bibinfo {author} {\bibfnamefont {M.}~\bibnamefont {Tamta}}, \bibinfo {author} {\bibfnamefont {N.}~\bibnamefont {Raj}}, \ and\ \bibinfo {author} {\bibfnamefont {P.}~\bibnamefont {Sharma}},\ }\href {\doibase 10.1103/PhysRevD.111.043043} {\bibfield  {journal} {\bibinfo  {journal} {Phys. Rev. D}\ }\textbf {\bibinfo {volume} {111}},\ \bibinfo {pages} {043043} (\bibinfo {year} {2025})},\ \Eprint {http://arxiv.org/abs/2405.20365} {arXiv:2405.20365 [astro-ph.HE]} \BibitemShut {NoStop}%
\bibitem [{\citenamefont {Graham}\ and\ \citenamefont {Ramani}(2024)}]{Graham:2023unf}%
  \BibitemOpen
  \bibfield  {author} {\bibinfo {author} {\bibfnamefont {P.~W.}\ \bibnamefont {Graham}}\ and\ \bibinfo {author} {\bibfnamefont {H.}~\bibnamefont {Ramani}},\ }\href {\doibase 10.1103/PhysRevD.110.075011} {\bibfield  {journal} {\bibinfo  {journal} {Phys. Rev. D}\ }\textbf {\bibinfo {volume} {110}},\ \bibinfo {pages} {075011} (\bibinfo {year} {2024})},\ \Eprint {http://arxiv.org/abs/2311.07654} {arXiv:2311.07654 [hep-ph]} \BibitemShut {NoStop}%
\bibitem [{\citenamefont {Yoo}\ \emph {et~al.}(2004)\citenamefont {Yoo}, \citenamefont {Chaname},\ and\ \citenamefont {Gould}}]{Yoo:2003fr}%
  \BibitemOpen
  \bibfield  {author} {\bibinfo {author} {\bibfnamefont {J.}~\bibnamefont {Yoo}}, \bibinfo {author} {\bibfnamefont {J.}~\bibnamefont {Chaname}}, \ and\ \bibinfo {author} {\bibfnamefont {A.}~\bibnamefont {Gould}},\ }\href {\doibase 10.1086/380562} {\bibfield  {journal} {\bibinfo  {journal} {Astrophys. J.}\ }\textbf {\bibinfo {volume} {601}},\ \bibinfo {pages} {311} (\bibinfo {year} {2004})},\ \Eprint {http://arxiv.org/abs/astro-ph/0307437} {arXiv:astro-ph/0307437} \BibitemShut {NoStop}%
\bibitem [{\citenamefont {Lu}\ \emph {et~al.}(2021)\citenamefont {Lu}, \citenamefont {Takhistov}, \citenamefont {Gelmini}, \citenamefont {Hayashi}, \citenamefont {Inoue},\ and\ \citenamefont {Kusenko}}]{Lu:2020bmd}%
  \BibitemOpen
  \bibfield  {author} {\bibinfo {author} {\bibfnamefont {P.}~\bibnamefont {Lu}}, \bibinfo {author} {\bibfnamefont {V.}~\bibnamefont {Takhistov}}, \bibinfo {author} {\bibfnamefont {G.~B.}\ \bibnamefont {Gelmini}}, \bibinfo {author} {\bibfnamefont {K.}~\bibnamefont {Hayashi}}, \bibinfo {author} {\bibfnamefont {Y.}~\bibnamefont {Inoue}}, \ and\ \bibinfo {author} {\bibfnamefont {A.}~\bibnamefont {Kusenko}},\ }\href {\doibase 10.3847/2041-8213/abdcb6} {\bibfield  {journal} {\bibinfo  {journal} {Astrophys. J. Lett.}\ }\textbf {\bibinfo {volume} {908}},\ \bibinfo {pages} {L23} (\bibinfo {year} {2021})},\ \Eprint {http://arxiv.org/abs/2007.02213} {arXiv:2007.02213 [astro-ph.CO]} \BibitemShut {NoStop}%
\bibitem [{\citenamefont {Inoue}\ and\ \citenamefont {Kusenko}(2017)}]{Inoue:2017csr}%
  \BibitemOpen
  \bibfield  {author} {\bibinfo {author} {\bibfnamefont {Y.}~\bibnamefont {Inoue}}\ and\ \bibinfo {author} {\bibfnamefont {A.}~\bibnamefont {Kusenko}},\ }\href {\doibase 10.1088/1475-7516/2017/10/034} {\bibfield  {journal} {\bibinfo  {journal} {JCAP}\ }\textbf {\bibinfo {volume} {10}},\ \bibinfo {pages} {034} (\bibinfo {year} {2017})},\ \Eprint {http://arxiv.org/abs/1705.00791} {arXiv:1705.00791 [astro-ph.CO]} \BibitemShut {NoStop}%
\bibitem [{\citenamefont {Manshanden}\ \emph {et~al.}(2019)\citenamefont {Manshanden}, \citenamefont {Gaggero}, \citenamefont {Bertone}, \citenamefont {Connors},\ and\ \citenamefont {Ricotti}}]{Manshanden:2018tze}%
  \BibitemOpen
  \bibfield  {author} {\bibinfo {author} {\bibfnamefont {J.}~\bibnamefont {Manshanden}}, \bibinfo {author} {\bibfnamefont {D.}~\bibnamefont {Gaggero}}, \bibinfo {author} {\bibfnamefont {G.}~\bibnamefont {Bertone}}, \bibinfo {author} {\bibfnamefont {R.~M.~T.}\ \bibnamefont {Connors}}, \ and\ \bibinfo {author} {\bibfnamefont {M.}~\bibnamefont {Ricotti}},\ }\href {\doibase 10.1088/1475-7516/2019/06/026} {\bibfield  {journal} {\bibinfo  {journal} {JCAP}\ }\textbf {\bibinfo {volume} {06}},\ \bibinfo {pages} {026} (\bibinfo {year} {2019})},\ \Eprint {http://arxiv.org/abs/1812.07967} {arXiv:1812.07967 [astro-ph.HE]} \BibitemShut {NoStop}%
\bibitem [{\citenamefont {Croon}\ and\ \citenamefont {Sevillano~Mu{\~n}oz}(2024)}]{Croon:2024rmw}%
  \BibitemOpen
  \bibfield  {author} {\bibinfo {author} {\bibfnamefont {D.}~\bibnamefont {Croon}}\ and\ \bibinfo {author} {\bibfnamefont {S.}~\bibnamefont {Sevillano~Mu{\~n}oz}},\ }\href {\doibase 10.1088/1475-7516/2024/07/060} {\bibfield  {journal} {\bibinfo  {journal} {JCAP}\ }\textbf {\bibinfo {volume} {07}},\ \bibinfo {pages} {060} (\bibinfo {year} {2024})},\ \Eprint {http://arxiv.org/abs/2403.13072} {arXiv:2403.13072 [astro-ph.CO]} \BibitemShut {NoStop}%
\end{thebibliography}%

\end{document}